\documentclass {article} 

\usepackage {apalike} 
\usepackage {amsfonts,graphics,amscd} 
\usepackage{undertilde}

\begin{document}   

\author{J. Brian Pitts \\  Faculty of Philosophy, University of Cambridge 
  \\ jbp25@cam.ac.uk }

\date{23 May 2014} 

\sloppy

\title{Change in Hamiltonian General Relativity from the Lack of a Time-like Killing Vector Field\footnote{Revised, accepted by \emph{Studies in History and Philosophy of Modern Physics}}}

\maketitle


\abstract{ In General Relativity in Hamiltonian form, change has seemed to be missing, defined only asymptotically, or otherwise obscured at best, because the Hamiltonian is a sum of first-class constraints and a boundary term and thus supposedly generates gauge transformations.
Attention to the gauge generator $G$ of Rosenfeld, Anderson, Bergmann, Castellani \emph{et al.}, a specially \emph{tuned sum} of first-class constraints, facilitates seeing that 
a solitary first-class constraint in fact generates not a gauge transformation, but a bad physical change  in electromagnetism (changing the electric field) or General Relativity.  The change spoils the Lagrangian constraints, Gauss's law or the Gauss-Codazzi relations describing embedding of space into space-time, in terms of the physically relevant velocities rather than auxiliary canonical momenta.  
While Maudlin and Healey have defended change in GR much as G. E. Moore resisted skepticism, there remains a need to exhibit the technical flaws in the no-change argument.

Insistence on Hamiltonian-Lagrangian equivalence, a theme emphasized by Mukunda, Castellani, Sugano, Pons, Salisbury, Shepley and Sundermeyer among others, holds the key. 
Taking objective change to be  ineliminable time dependence, one recalls that there is change in vacuum GR just in case there is no time-like vector field $\xi^{\alpha}$ satisfying Killing's equation $\pounds_{\xi}g_{\mu\nu}=0$, because then there exists no coordinate system such that everything is independent of time.
   Throwing away the spatial dependence of GR for convenience, one finds explicitly that the time evolution from Hamilton's equations is real change  just when there is no time-like Killing vector.  The inclusion of a massive scalar field is simple. No obstruction is expected in including spatial dependence and coupling more general matter fields. Hence change is real and local even in the Hamiltonian formalism. 

The considerations here resolve the Earman-Maudlin standoff over change in Hamiltonian General Relativity:  the Hamiltonian formalism is helpful, and, suitably reformed, it does not have absurd consequences for change. 
Hence the classical problem of time is resolved, apart from the issue of observables, for which the solution is outlined.  The Lagrangian-equivalent Hamiltonian analysis of change in General Relativity is compared to Belot and Earman's treatment.   The more serious quantum problem of time, however, is not automatically resolved due to issues of quantum constraint imposition.  

}

 Keywords: constrained Hamiltonian dynamics, General Relativity, problem of time, quantum gravity, variational principles

\tableofcontents

\section{Introduction}

\subsection{Hamiltonian Change Seems Missing but Lagrangian Change Is Not}  

It has been argued  that General Relativity, at least in Hamiltonian form, lacks change, has change only asymptotically and hence only for certain topologies, or appears to lack change with no clear answer in sight (\emph{e.g.}, \cite{AndersonChange,IshamTime,BelotEarman,EarmanMcTaggart,RicklesTimeStructureQG,HuggettWuthrichTimeQG}).   Such a conclusion calls to mind earlier philosophical puzzles, whether ancient (the paradoxes of Zeno, whom James Anderson mentions \cite{AndersonRoyaumont,AndersonChange}, and the views of Parmenides, whom Kucha\v{r} mentions \cite{KucharCanonical93}) or modern (the argument concluding that real time requires something contradictory and hence is impossible by McTaggart \cite{McTaggart}, mentioned in a memorable philosophical exchange \cite{EarmanMcTaggart,MaudlinMcTaggart}).  The new conclusion, following apparently with mathematical rigor from our standard theory of gravity, is not as readily ignored as Zeno, Parmenides and McTaggart. On the other hand, if one breathes the fresh, clean air of numerical relativity from time to time, it is difficult not to notice that there really is change in GR.  Thus one might suspect  that any formal Hamiltonian results to the contrary are mistaken. 
Such a conclusion is all the clearer if one recalls that the Lagrangian GR formalism and 4-dimensional differential geometry are not thought to have any analogous problem.   Either the canonical standards are inappropriately strict, or the 4-dimensional Lagrangian standards are too loose.  But no one thinks the latter.

The Earman-Maudlin philosophical exchange provides a good starting point \cite{EarmanMcTaggart,MaudlinMcTaggart}.  Maudlin displays liberal amounts of common sense about point individuation, observables (in the non-technical sense of what can be observed), \emph{etc.}, whereas Earman displays standard glosses on standard mathematical physics.  Neither common sense rooted in scientific practice nor a common interpretation of mathematical physics is to be taken lightly, but can one have both?
Is there a point in which mathematical physics becomes so bizarre as to undermine itself by excluding grounds for any possible empirical confirmation \cite{HealeyGRchangelessincoherent}?  Earman has elsewhere composed an ``Ode'' commending the Dirac-Bergmann constrained dynamics formalism to philosophers \cite{EarmanOde}.  The reader of the Earman-Maudlin exchange gets the impression that  each side declares victory.
The progress of physics has been so great, and often enough counterintuitive, that beating back Poisson brackets  with appeals to common sense  does not yield full conviction, and rightly so.    If Earman unwittingly exhibits ``How to Abuse Gauge Freedom to Generate Metaphysical Monstrosities,'' as Maudlin's subtitle claims, then what is the right way to handle gauge freedom? On this question Maudlin is less full than one would prefer.  
Ultimately I will side with Maudlin's common-sense conclusions, though not his dismissive view of the Hamiltonian formalism. Change will be defended not in defiance of or indifference to mathematical physics, but through careful engagement in it and reform  motivated by more solid mathematical physics---in line with Maudlin's invocation of the gold standard formulation of GR in terms of Einstein's equations and 4-dimensional differential geometry.


\subsection{Maudlin's and Healey's Critiques in G. E. Moore's Style}

One could  affirm real change in GR without attending at all to arguments about the Hamiltonian formalism, because nothing about the Hamiltonian formalism's treatment of change could be more decisive  than the meaning of the presence or absence of a time-like Killing vector.  
This claim bears a  resemblance to the response to skepticism by G. E. Moore \cite{GEMooreExternalWorld}, as well as the spirit of Maudlin's and Healey's responses to Earman \cite{MaudlinMcTaggart,HealeyGRchangelessincoherent}.
 But the Moorean-like fact, in my view, is not (or not only) some deliverance of common sense, accessible by simple bodily gestures (Moore's displaying his hands, Samuel Johnson's kicking a stone), but rather (or also), a deliverance of Lagrangian field theory. 
Yet this is no justification for dismissing the Hamiltonian formalism. 
 It is, rather, a  call  for reform.  

The necessity and incompleteness of such an approach resembles Norman Malcolm's discussion of Moore's philosophy\footnote{I owe this reference to Jim Weatherall.}:
\begin{quote}
Two things may be said against Moore's method of refutation.  (Footnote:  This must be taken as qualifying my previous statement that Moore's refutations are \emph{good} ones.)  In the first place, it often fails to convince the author
of the paradox that he is wrong.\ldots
In the second place, Moore's style of refutation does not get
at the sources of the philosophical troubles which produce the
paradoxes.\ldots
Although Moore's philosophical method is an incomplete
method, it is the essential first step in a complete method. The
way to treat a philosophical paradox is first of all to resist it,
to prove it false. Because, if the philosopher is pleased with his
paradox, fancies it to be true, then you can do nothing with
him. It is only when he is dissatisfied with his paradox, feels
refuted, that it is possible to clear up for him the philosophical
problem of which his paradox is a manifestation.
\cite[pp. 366-367]{MooreSchilppMalcolm} \end{quote}
If Malcolm's praise seems too strong (because Moorean anti-skeptical arguments are not always necessary and not always good), the point remains that Moore-style arguments such as Maudlin's and Healey's are sometimes good, incomplete, and yet inspirational unraveling a flawed skeptical argument, as is the case here.  The considerations presented above largely fill the gap left by the Moorean style of defense of change in GR.

\section{Lagrangian Interpretive Strategy Brings Clarity}

Attending to the Lagrangian formalism of General Relativity and to 4-dimensional differential geometry  holds the key to clarity in all these matters. It seems to be widely agreed on diverse grounds that the Lagrangian formulation of mechanics (broadly construed) is more fundamental than the Hamiltonian one  
 \cite{GotayNesterPoincare,CurielLagrangian}.\footnote{There is, to be sure, a Hamiltonian derivation of geometrodynamics \cite{HojmanKucharTeitelboim}.  Whether one can seriously imagine someone first finding GR by that means is another matter.  One risk of a freestanding Hamiltonian view is the temptation (resisted by these authors but not others) to forget that one only learns what the canonical momenta mean physically by virtue of the equations $\dot{q}=\frac{\delta H}{\delta p}.$ By contrast the Lagrangian lacks those \emph{a priori} physically meaningless dynamical quantities.  That is one clear respect in which the Lagrangian formalism is more fundamental than the Hamiltonian.  For such reasons, it is best to direct one's thoughts to the Hamiltonian action $\int dt (p \dot{q}-H)$ rather than the Hamiltonian itself. } It is also widely believed that  the two are equivalent (apart perhaps from topological restrictions), or at least that they should be.  Yet there are controversies in the literature on constrained dynamics about whether such equivalence actually holds, and various proofs presented have too narrow a scope (such as addressing the equivalence of equations of motion but neglecting to address the equivalence of the gauge transformations).   
 One possible view is described 
by Pons, Salisbury and Sundermeyer \cite{PonsSalisburySundermeyerFolklore}:
\begin{quote}
[t]he position on one side is that there ought to be no debate at all [about the physical interpretation of General Relativity or any generally covariant theory] because the phase space formalism is equivalent to the formalism in configuration-velocity space, and no one has claimed that any interpretational problem exists in the latter framework.  Entire books have been devoted to the experimental tests of GR, and this very language implies that observables exist - alive and kicking.  Thus the entire debate must be a consequence of misunderstandings. \cite[p. 3]{PonsSalisburySundermeyerFolklore}.  
\end{quote}
Such a view suggests a Lagrangian-first interpretive strategy.

This view does  not seem to be the view of Kucha\v{r}, though he, very unusually, is willing to tinker with the Dirac-Bergmann formalism to uncover real change in General Relativity  \cite{KucharCanonical93}.  Kucha\v{r}'s reinterpretation of the Hamiltonian constraint  is not  systematic---the common-sense arguments about observing temporal change work equally well for the momentum constraint and spatial change. 
 Neither is Kucha\v{r}'s view  clearly inspired by the need for equivalence with the  Lagrangian formalism. 
He  allows that observables should commute with the momentum constraint $\mathcal{H}_i,$ because we cannot directly observe spatial points.  But he denies that observables should commute with the Hamiltonian constraint $\mathcal{H}_0.$  He notes that one cannot directly observe which hypersurface one is on, either---which provides pressure to think that observables should commute with the Hamiltonian constraint.  But he does not treat the constraints even-handedly despite their fitting his argument-form equally well:
\begin{quote} However, the collection of the canonical data $g_{(1)}$, $p_{(1)}$ on the
first hypersurface is clearly distinguishable from the collection $g_{(2)}$, $p_{(2)}$ of the evolved
data on the second hypersurface. If we could not distinguish those two sets of the data,
we would never be able to observe dynamical evolution. \cite[p. 137]{KucharCanonical93} \end{quote} 
 Indeed so.  But who makes observations over an entire space-like hypersurface?  We observe that the world is spatially varying, so, by parity of reasoning, why should observables commute with the momentum constraint $\mathcal{H}_i,$ either?  Hence making his argument more consistent provides a reason to doubt that observables should commute with the Hamiltonian or momentum constraints.  One could also note that neither $H_0$ by itself nor $H_i$ actually generates a gauge transformation due to neglect of the lapse and shift vector, basically the part of the space-time metric that is not contained in the spatial metric.

 Kucha\v{r}'s common-sense remarks on what experimental physicists routinely observe and Maudlin's common-sense remarks about observables point in the same direction. But the view described isn't quite that view of  Maudlin, either \cite{MaudlinMcTaggart}.  While Maudlin is satisfied with the Lagrangian formalism and its implicit concept of observables in terms of tensor calculus, he takes the problems in Hamiltonian  General Relativity as a reason to reject it, not to reform it.  Such a response is both unnecessarily drastic and incompatible with regarding the phase space formalism as equivalent to the configuration-velocity formalism.

The view described above by Pons, Salisbury and Sundermeyer \cite{PonsSalisburySundermeyerFolklore} is  attractive, if suitably developed to reform the Dirac-Bergmann formalism to achieve Hamiltonian-Lagrangian equivalence, and is adopted here.  
 A principled way to implement the common sense that Kucha\v{r} and Maudlin forcefully deploy is to apply consistently the standards of observability/physical reality in Lagrangian/4-dimensional GR, where tensor fields (if not hobbled by other gauge freedoms like electromagnetism) are observable.  
 Something like this view is implicit in the Pons-Salisbury-Shepley-Sundermeyer principles of reforming the Dirac-Bergmann formalism to ensure Hamiltonian-Lagrangian equivalence and paying due attention to the gauge generator $G$ \cite{AndersonBergmann,CastellaniGaugeGenerator}.  
One can take as a motto a remark of Pons and Shepley that has characterized that series of works (\emph{e.g.}, 
   \cite{PonsSalisburyShepley,ShepleyPonsSalisburyTurkish,PonsSalisbury,PonsReduce}):
\begin{quote}
We have been guided by the principle that the Lagrangian and Hamiltonian formalisms should be equivalent\ldots
in coming to the conclusion that they in fact are.  \cite[p. 17]{PonsReduce}
\end{quote}
Indeed one can hardly do otherwise in constrained dynamics without falling into inconsistency or error.  But such an innocuous principle, they have shown, has unappreciated consequences that yield greater conceptual clarity.  In this paper I aim to push further, primarily on conceptual rather than technical matters, in the same direction.
The descriptive claim that the $q$-$p$ formalism is equivalent to the $q$-$\dot{q}$ formalism is too quick; that equivalence \emph{should} hold, but work has been needed to make it hold.

\subsection{Taproot of Confusion:  First-Class Constraints and Gauge Transformations}

Indeed still more work is needed to make the $q$-$p$ formalism equivalent to the $q$-$\dot{q}$ formalism.  Recently I showed that if one accepts the longstanding claim that a first-class constraint generates a gauge transformation, then one spoils Gauss's law, the Lagrangian $q$-$\dot{q}$ constraint $\nabla \cdot \vec{E}=0$ in (vacuum) electromagnetism \cite{FirstClassNotGaugeEM}. 
The electric field, a familiar function of the derivatives of the potentials $A_{\mu}$ \cite{Jackson}, changes by the gradient of the arbitrary smearing function and hence ceases to be divergenceless even in the absence of charge.  It is easy not to notice this problem because the conjugate momentum that one expected to be the electric field, remains divergenceless. If one isn't careful to retain Hamiltonian-Lagrangian equivalence, the conjugate momenta can cease to mean what one expected, as Anderson and Bergmann pointed out long ago \cite{AndersonBergmann}.

To see the problem, one can add the two independently smeared constraints' actions together:
\begin{eqnarray} \delta A_{\mu}(x) = \{ A_{\mu}(x), \int d^3y [p^0(y) \xi(t,y)      + p^i,_i(y) \epsilon(t,y)] \}= \delta^0_{\mu} \xi  -\delta_{\mu}^i \partial_i \epsilon. \end{eqnarray}   
Then one calculates the 4-dimensional curl of the transformed 4-vector potential less the curl of the original potential, getting the curl of the gauge transformation due to linearity. (The fact that the electric field itself is not defined on phase space is not relevant.  Once one knows $A_{\mu},$ one can take its curl to find the electromagnetic field.)  The magnetic field is unchanged \cite[p. 134]{Sundermeyer}.  But  the electric field, which curiously has been neglected, \emph{does} change  \cite{FirstClassNotGaugeEM}---a result that dashes  expectations that a first-class constraint alone generate a gauge transformation.  The combined change in $\vec{E}$ is given by
\begin{eqnarray} \delta F_{0n } = -\delta \vec{E} =  - \partial_{n}\xi  -   \partial_n \partial_{0} \epsilon.  \end{eqnarray}
If one puts the constraints to work together as a \emph{team} by setting $\xi=-\dot{\epsilon}$ to make the $\delta F_{0n}=0,$  then   
\begin{eqnarray} \delta A_{\mu}(x) = \{ A_{\mu}(x), \int d^3y [-p^0(y) \dot{\epsilon}(t,y)      + p^i,_i(y) \epsilon(t,y)] \} =  -  \delta^0_{\mu} \dot{\epsilon}  -\delta_{\mu}^i \partial_i \epsilon = -  \partial_{\mu} \epsilon, \end{eqnarray}  which is good. 
  Not surprisingly in light of the form of the gauge generator  \cite{AndersonBergmann,CastellaniGaugeGenerator,PonsSalisburyShepleyYang} \begin{eqnarray} G=\int d^3x (p^i,_i \epsilon - p^0 \dot{\epsilon}),\end{eqnarray} $p^0$ and $p^i,_i$  generate \emph{compensating} changes in $\vec{E}$  when suitably combined.
Indeed one can piece together $G$  by demanding that the changes in $\vec{E}$ cancel out.  Two wrongs, with opposite signs and time differentiation, make a right.  This tuning, not surprisingly, is a special case of what Sundermeyer found necessary to get first-class transformations to combine suitably to get the familiar gauge transformation for the potentials for Yang-Mills \cite[p. 168]{Sundermeyer}. The benefit of my taking the curl \emph{before} fixing the coefficients of the constraints to make a team is  to show that getting the usual Lagrangian gauge transformations is compulsory on pain of spoiling elementary facts of electromagnetism, not merely optional and  comfortingly familiar  as it has seemed to be in the constrained dynamics literature.  The commutative diagram makes the point.
$$\begin{CD}
A_{\mu}     @>L-equiv.>>  G=\int d^3x(-p^0\dot{\epsilon}+\epsilon p^i,_i) @>>> \delta A_{\mu}= -\partial_{\mu} \epsilon  \\
@V\int d^3x(p^0\xi+\epsilon p^i,_i)VV                      @.                      @VVcurlV \\
 \delta A_{\mu} = \delta^0_{\mu} \xi - \delta^i_{\mu} \epsilon,_i  @>curl>>   \delta F_{\mu\nu}  = (\delta^0_{\nu}  \xi,_{\mu}   - \delta^i_{\nu}  \epsilon,_{i\mu}) - \mu \leftrightarrow \nu   @>L-equiv.>\xi=-\dot{\epsilon}>  \delta F_{\mu\nu}=0
\end{CD}$$
While the top line is not unknown in works advocating  Hamiltonian-Lagrangian equivalence, the bottom line appears to be novel.  
It is of course unacceptable to have $\delta F_{\mu\nu} \neq 0,$ so requiring  Lagrangian equivalence from the Hamiltonian resolves the trouble. Hamiltonian-Lagrangian equivalence is the law, not just a good idea.

The equality (up to a sign) of the electric field and the momentum $p^i$ conjugate to $A_i$ \emph{using Hamilton's equations}  $$\dot{A}_{i}=\frac{\delta H}{\delta p^i }$$ has tempted  many authors to treat that equality as if it were something stronger.  These are among the boring Hamilton's equations, ones that recover what one already knew in the Lagrangian context and then forgot in performing the Legendre transformation.  Boring or not, that relationship must be forgotten in calculating Poisson brackets.  Some authors, far from forgetting that equality, even use the letter  $E$  for the canonical momentum, reflecting in notation an overly hasty identification that is often less transparently made  \cite{FaddeevEnergy,BelotEarman}.  But that canonical  momentum is no longer equivalent to the electric field after either first-class constraint has acted on $A_{\mu}.$  Likewise, one spoils the Hamiltonian and momentum constraints in GR---at least the physically relevant versions, which are in terms of  $q$ and $\dot{q}$ (the 3-metric, lapse, shift vector, and extrinsic curvature tensor, roughly) and pertain to intrinsic curvature and the embedding of space into space-time, not $q$ and off-shell physically meaningless canonical momenta $\pi^{ij}$  by acting with a primary first-class constraint or a secondary first-class constraint.   Hence a first-class constraint in these representative theories, whether primary or secondary, generates a bad physical change, not a gauge transformation. 
Conjugate momenta don't ultimately matter; they (at least the unconstrained ones) are just auxiliary fields for making $\dot{q}$ (and hence $q$) do the right thing over time. Being auxiliary fields, they appear essentially algebraically in the canonical action $$\int dt d^3x (p \dot{q} - \mathcal{H}) $$     and can be `eliminated' (reduced to functions of derivatives of $A_{\mu}$ in physically possible worlds) using their own equations of motion, leading back to the original Lagrangian action $\int dt d^3x \mathcal{L}$.  Such an ontologically superfluous entity clear is not the primordial observable electric field.  
 In electromagnetism, coupling to charge happens through an interaction term $A_{\mu} J^{\mu};$ $p^i$ is nowhere in sight.  
If the $q$'s---in this case $A_{\mu}$---do the right thing over time, all is well; if they don't (no matter how well the conjugate momenta behave), all is lost.  In the context of GR, if the momentum $\pi^{ij}$ conjugate to the 3-metric $h_{ij}$ behaves properly, but the extrinsic curvature tensor $K_{ij}$ behaves improperly, then space doesn't fit rightly into space-time and Einstein's equations are false.  Usually one doesn't bother with this distinction because the boring Hamilton's equations $\dot{h}_{ij} = \frac{\delta H}{\delta \pi^{ij}}$ return the correct relationship that one forgets in setting free $ \pi^{ij}$ in the Legendre transformation.  But that relationship is not an identity. It is  just a field equation (`on-shell'), whereas gauge transformations are (typically, including electromagnetism and the momentum constraint $\mathcal{H}_i$ in GR) defined off-shell.  Hence an inept identification of gauge transformations can spoil $\dot{h}_{ij} = \frac{\delta H}{\delta \pi^{ij}}$ or its electromagnetic analog, making the momenta lose their usual relationship to the velocities.   $\dot{h}_{ij} = \frac{\delta H}{\delta \pi^{ij}}$ arrives on the scene logically too late to prevent trouble; instead it just makes a mess.
 Hence good behavior of the momenta does not guarantee good behavior of the velocities.  The latter can, and should, be checked directly. Naturally there will be agreement in a correct Hamiltonian formalism.  With such issues in mind Anderson and Bergmann urged attending to the Lagrangian constraint surface, not only the Hamiltonian constraint surface \cite{AndersonBergmann}.

After 30 years the force of Castellani's recovered insight about the team rather than individual character of gauge transformation generation by first-class constraints  still has not been generally recognized.  He  said that 
\begin{quote}
Dirac's conjecture that all secondary first-class constraints generate
symmetries is revisited and replaced by a theorem.\ldots  The old question whether secondary first-class constraints generate gauge
symmetries or not \ldots is then solved: they are \emph{part} of a gauge generator
$G$ \ldots \cite[pp. 357, 358]{CastellaniGaugeGenerator}. (emphasis in the original)  
\end{quote}
The force of ``replaced'' requires  the \emph{elimination of the old erroneous claim}, not just the introduction of a new true claim. Nowadays one sees a curious coexistence of beliefs in first class constraints as generating gauge transformations (or lingering consequences of that old belief pertaining to change or observables) and belief in the gauge generator $G$, a special \emph{combination} of first class constraints, as generating gauge transformations.  
 Indeed the gauge generator is already a very old part of the constrained dynamics literature \cite{AndersonBergmann,RosenfeldQG,SalisburyRosenfeldMed}, so in one sense there wasn't much of a new claim to introduce even in 1982 (apart from the use of a $3+1$ split, the 1958 trivialization of the primaries by a well chosen boundary term \cite{DiracHamGR,AndersonPrimary}, and elimination of $\dot{q}$ from the gauge generator in favor of Legendre projectability)---though the gauge generator had been almost entirely forgotten apart from recent work by Mukunda \cite{MukundaGaugeGenerator}. 
In a sense the Dirac conjecture is two-sided:  it claims that every first-class constraint is involved in generating gauge transformations and that each first-class constraint that generates a gauge transformation does so  by itself. (Actually the conjecture pertains only to secondary and higher generations of constraints, because Dirac thought that he had proved the result for primary constraints \cite[p. 21]{DiracLQM}.) The former claim, which is generally received, seems to be true apart from ineffective constraints (such as squared constraints, which have vanishing gradient and hence vanishing Poisson brackets on the constraint surface).  This former claim seems to account for some recent literature that purports to prove the Dirac conjecture while in fact arriving at the gauge generator $G$ instead \cite{GitmanTyutinGauge}.    The latter claim is false.

While Castellani's attention (following Dirac) was initially directed to secondary constraints, it is easy to show by direct calculation that a \emph{primary} constraint in Maxwell's electromagnetism (or in General Relativity) does not generate a gauge transformation either.  In electromagnetism it changes the electric field $\vec{E}$ by a gradient, just as the secondary constraint does \cite{FirstClassNotGaugeEM}, again spoiling Gauss's law.  Not surprisingly, a special combination of the two leaves the electric field (and the magnetic field, which the constraints do not touch) unchanged. That combination is just the gauge generator   \cite{AndersonBergmann,CastellaniGaugeGenerator,PonsSalisburyShepleyYang} 
 $$G=\int d^3x (p^i,_i \xi - p^0 \dot{\xi}).$$ Dropping a boundary term gives the more suggestive form $$G=-\int d^3x p^{\nu} \partial_{\nu} \xi.$$  $G$ acts as $$\{ G, A_{\mu} \}= \partial_{\mu} \xi $$ and  $$\{ G, p^{\mu} \} =0.$$  Hence one can reinvent the gauge generator in electromagnetism simply by requiring that the changes in the electric field due to the first-class constraints cancel out. One can revisit Dirac's argument that a first-class constraint generates a gauge transformation \cite[p. 21]{DiracLQM} with electromagnetism in mind.  
Part of the  problem is that Dirac, by comparing two solutions with identical initial data, neglects the effect that a gauge transformation can have instantaneously on the initial data surface, thus underestimating the influence of the primary constraints.  Relatedly, he neglects the fact that the secondary first-class constraint (the one like Gauss's law) appears in the Hamiltonian with a gauge-dependent coefficient $-A_0$, so two evolutions will (eventually) differ not simply by the coefficients of the primary constraint (the scalar potential's velocity), but also by what the Gauss-like constraint generates \cite{FirstClassNotGaugeEM,PonsDirac}.  To see the difference, notice the effects on the vector potential $A_i,$ which is certainly altered (by a gradient) by a gauge transformation, but is not altered by the primary constraint.  
 Unfortunately Dirac's error is widely followed \cite{Govaerts,HenneauxTeitelboim,WipfBadHonnef,RotheRothe}. (By contrast Sundermeyer computes relevant Poisson brackets providing the raw material to diagnose the problem \cite[p. 134]{Sundermeyer}.) This oversight of Dirac's motivates the extended Hamiltonian formalism, which is intended to recover the full gauge freedom which is actually already there in the primary Hamiltonian with the primary and secondary first-class constraints.

 For General Relativity, there is an additional problem in Dirac's argument regarding the primary constraint conjugate to the lapse $N$  \cite{BarbourFosterPrimary,ThebaultReductionProblemofTime}.  The reason is that  his argument 
assumes that physically equivalent points should have equal values of the coordinate time---in effect, that the  time is observable, as was noticed  by Henneaux and Teitelboim \cite[pp. 17-19]{HenneauxTeitelboim}.  They  attempt to fill the gap another way \cite[p. 107]{HenneauxTeitelboim}, but give up Hamiltonian-Lagrangian equivalence in the process.

This paper continues the project of resolving problems in canonical General Relativity by enforcing the equivalence of the (obscure and troubled) Hamiltonian formalism to the (perspicuous and correct) Lagrangian formalism.  The clear formalism interprets the unclear one.  
Maudlin's common sense about change and observables is inspirational, but his disregard for the Hamiltonian formalism  is not satisfactory.
  What is needed is  
conceptual-technical work that identifies distinctively Hamiltonian conjectures about gauge freedom, change, and observables 
and eliminates them in favor of Lagrangian-equivalent notions. Paradoxical features of the Hamiltonian formalism are
a good place to look for such problems.  Sometimes one can show that seemingly paradoxical features are equivalent to the plainly correct Lagrangian or differential geometric concepts and hence, on further reflection, no longer paradoxical. In other cases paradoxical Hamiltonian claims are erroneous.  
  The predecessor paper \cite{FirstClassNotGaugeEM} carried out that task for gauge freedom and first class constraints for electromagnetism. As a consequence one wants observables to be defined in terms of suitable Poisson brackets not with first-class constraints separately, but with the gauge generator $G$ \cite{PonsSalisburySundermeyerFolklore}. 
  Another paper (in preparation) carries out the same technical task for General Relativity \cite{FirstClassNotGaugeGR}.  This current paper carries out the task regarding change in GR using a simplification, discarding spatial dependence. Hence most of the relevant conceptual features will be found (apart from many-fingered time), while avoiding much irrelevant technical complication.    A following paper will carry out the task for observables in GR, where it turns out that there is a key distinction between internal and external symmetries; as a result, Hamiltonian-Lagrangian equivalence for observables involves not \emph{vanishing} Poisson bracket (with $G$), but a Poisson bracket with $G$ that gives the Lie derivative of a geometric object. Meditation on the role of the transport term in the Lie derivative shows why the absurd result of constant observables arises from imposing the vanishing of the Lie derivative of observables with respect to an arbitrary vector field describing the gauge transformation: one might as reasonably ask for observable features of the world to be the same at 1 am Eastern Daylight Time and 1 am Eastern Standard Time an hour later.   While the ideas about change and observables can be understood in isolation, the overall coherence of the package is best appreciated collectively.  Clearing away confusion about first-class constraints clears the field of entrenched errors about change and observables to make the Lagrangian-equivalent definitions more plausible.


\subsection{Invariance of Action \emph{vs.} First-Class Constraints}

The argument above that either first-class constraint in electromagnetism is a bad physical change assumes that the electromagnetic field $F_{\mu\nu}$ should count as observable.  One can consider other theories in which the observables are unknown, are not local fields (as in Yang-Mills), or are controversial.  A proper Hamiltonian formalism ought itself to specify what the observables are.  Hence a perhaps even more compelling argument that a first-class constraint does not generate a gauge transformation would refer not to observables, but to the action principle:  a gauge transformation is a time- and (typically) space-dependent redescription that changes the action by at most a boundary term (equivalently, changes the Lagrangian density by at most a divergence).  One can show for the canonical action that each first-class constraint generates a change that isn't at most a boundary term and hence is not a gauge transformation, while the specially tuned sum $G$, the gauge generator, does change the action at most by a boundary term and hence is a gauge transformation \cite{FirstClassNotGaugeEM}.  
Hence one does not need to know in advance what the observables are or even what one means by observables in order to test a change of $q$ and $p$ for being a gauge transformation.  
Below the analogous argument will be exhibited for a simplified relative of GR.

\subsection{Illustration \emph{via} Homogeneous Truncation of GR}

 This paper aims to deploy just enough technical apparatus to make the point about change clear---that is, warranted and also not buried in irrelevancies---using a reparametrization-invariant `mechanical' theory derived by simply dropping the spatial dependence from GR.  
Such work sometimes goes by the name ``minisuperspace''; the particularly simple case addressed here corresponds to Bianchi I cosmological models.  My  aim is not primarily to illuminate a very narrow sector of GR (though in many respects the toy theory does so), but to have a `mechanical' system that is analogous to GR in most ways relevant to the task at hand, yet allows easy calculations.  
It has a physically undetermined function relating physical time to coordinate label time. This function, the ``lapse function'' $N$, appears undifferentiated in the Lagrangian and the Hamiltonian.  It relates to the space-time metric as $g^{00} =-N^{-2}$ \cite{MTW}.  The constrained Hamiltonian formalism due to Dirac, Bergmann, \emph{et al.} associates  with  $N$ a conjugate momentum $p,$ which is appended linearly to the Hamiltonian with an arbitrary function $v(t),$ giving the ``primary'' (or sometimes ``total,'' but some authors use the term differently \cite{Govaerts}) Hamiltonian  $H_p = N \mathcal{H}_0 + v p$, which generates time evolution \cite{Sundermeyer}.  Whether this time evolution constitutes objective change will be explored.
The distinction between this Hamiltonian (generator of time evolution from initial data) and the gauge generator $G$ will be made, used and discussed in detail. What will be achieved is a positive and intelligible Hamiltonian story about change that replaces the mysterious one built on the assumption that a first-class constraint generates a gauge transformation.

My treatment differs from Pons and Shepley's treatment of homogeneous cosmologies  \cite{PonsReduce} primarily due to a different aim.  
My aim is technically much simpler, namely, to find a theory that has enough GR-like behavior to threaten to lose time evolution just as GR does---time reparametrization invariance with a Hamiltonian constraint quadratic in the momenta---but simple enough that explicitly finding time evolution is easy when one looks in the right place. 
Confusingly, the counting of degrees of freedom in Bianchi cosmologies depends on the intended interpretation and also the global topology \cite{PonsReduce,AshtekarBianchi,SalisburyBianchiI}.  If one views such cosmologies as a sector of General Relativity (without preferred simultaneity), then solutions related by a 4-dimensional coordinate transformation will count as equivalent.  Thus one achieves the same number of degrees of freedom in the Hamiltonian and Lagrangian formalisms, as one should, if one is careful and insistent on doing so \cite{PonsReduce}.   While the Lagrangian and Hamiltonian treatments of the `mechanical' system ought to agree with each other,  for my conceptual purposes, neither many-fingered time nor the momentum constraints and their first-class character, are deliberately retained, sought, or valued. (I will have something to say about spatial variation and many-fingered time below; I do not anticipate any essential complication or novelty compared to what happens here in treating reparametrization-invariance.)  There will be no occasion to worry about homogeneity-preserving diffeomorphisms or the number of degrees of freedom.  Leaving redundant components in the spatial metric is useful in order to preserve the visual-mathematical resemblance between a tensor (matrix) field in GR and a matrix, as opposed to whittling it down to its diagonal elements (\emph{c.f.}, \cite{GoldbergKlotz,RyanShepley,SalisburyBianchiI}).  
 My treatment also assumes the simplest version of homogeneity, namely, with the three spatial Killing vector fields all commuting (vanishing structure constants, Bianchi type I), so one can simply keep $x^0=t$ and drop $x^i$ ($i=1,2,3$). Simultaneity is absolute, but the labels of the slices are not quantitatively meaningful.  
I expect that the treatment of change in terms of the lack of a time-like Killing vector field would hold for GR just as for the mechanical theory, but with tedious and irrelevant calculations tending to obscure the main point.  It will be necessary to do a bit of logic, however, paying attention to or's ($\lor$), and's ($\&$), not's ($\lnot$), and existential and universal quantifiers ($\exists$) and ($\forall$),  in order to show the equivalence of the Hamiltonian and differential geometric conceptions of change.  Some classical differential geometry, especially involving scalar densities and their Lie derivatives \cite{Schouten,Anderson}, will be needed in making the $3+1$ split of the space-time Killing vector equation into space (largely trivial) and time (the focus here).


\section{Change Unproblematic in 4-Dimensional  Geometry}

\subsection{Hamiltonian Impersonators  Obscure Change Generator}

If one attempts to define real change in terms of what some  Hamiltonian-like entity generates, one has to try to figure out which  Hamiltonian-like entity to use.  Even then one might not be sure that even one's favorite candidate generates real change.  Is it the Hamiltonian constraint $\mathcal{H}_0$ (also called $\mathcal{H}_{\perp}$)? 
One might find this line from Sundermeyer's book tempting: ``[s]o we really infer that $\mathcal{H}_{\perp}$ is responsible for the dynamics.'' \cite[p. 241]{Sundermeyer}  Is it perhaps instead the weight $2$ Hamiltonian constraint $\sqrt{h}\mathcal{H}_0,$ which avoids
awkward powers of the square root of the determinant of the metric and relates more closely to hyperbolic formulations of the field equations  \cite{YorkHamilton}?  (Indeed the former property can be achieved also for weight $4$, weight $6,$ weight $8,$ \emph{etc.} by throwing more powers of $h$ onto $\sqrt{h}\mathcal{H}_0$ and compensating with an oppositely weighted lapse, now taken  as primitive.)
Is it the  canonical Hamiltonian $H_c=N\mathcal{H}_0$?  It has the virtue of being what results from $p \dot{q} -L$ once the primary constraint is used to annihilate the coefficient of $\dot{N}$.   Is it  the primary Hamiltonian  $H_p = N \mathcal{H}_0 + v p$, which adds in the primary constraint(s)?\footnote{If one is  willing to keep $\dot{N}$ in the Hamiltonian formalism, then one can use the Sudarshan-Mukunda Hamiltonian instead  \cite[p. 93]{SudarshanMukunda} \cite{CastellaniGaugeGenerator}. } 
Is it the gauge generator $G,$ which is built out of the same secondary and primary constraints as the primary Hamiltonian, and which appears to include the primary Hamiltonian as a special case? Is it the extended Hamiltonian, which adds the secondary constraints by hand, as Dirac invented to try to find the most general motion possible \cite{DiracLQM}? Dirac's  attitude about whether any particular Hamiltonian was the right one was quite relaxed.\footnote{I thank Edward Anderson for a useful comment on this point.}  But one might expect that at most one Hamiltonian would yield real change equivalent to the reliable Lagrangian/differential geometric definition.  
With at least six moderately plausible candidates, it is best to look elsewhere besides the Hamiltonian formalism for some more reliable criterion to decide which, if any, generates real change.

Whether there really is change in GR cannot depend on whether the theory is described in terms of the Lagrangian or the Hamiltonian formalism.   But since the answer is a clear ``yes'' for the Lagrangian case, while the Hamiltonian formalism has long been obscure on the matter, it follows that the Hamiltonian answer needs to be ``yes'' as well.  If some Hamiltonian formalism does not give a positive answer, then one needs to rethink the formalism until a positive answer and equivalence to the Lagrangian formalism are achieved. 
Thus we will find that there is an answer that generates real change, and which of the six (or more) candidates it is.

\subsection{Time Dependence from Lack of Time-like Killing Vector Field }

It is best to get back to basics---something even more basic than the Lagrangian formalism, in fact.  Change, one might say, is being different at different times.  At least that definition would be adequate if General Relativity didn't pose the risk of fake change due to funny labeling.  A revised, more GR-aware definition would be that change is being different at different times in a way that is not an artifact of funny labeling.  A solution of Einstein's equations displays objective change iff (and where) it depends on time for all possible time coordinates.  At this point differential geometry comes in.

General Relativity expressed in the language of Lagrangian field theory and four-dimensional differential geometry is generally not believed to suffer from a problem of time or a lack of observable quantities  
 \cite{PonsSalisburySundermeyerFolklore}.  One of course needs to account for the coordinate (gauge) freedom, but that is not difficult to do---tensors and all that. (All transformations are interpreted \emph{passively}, thereby averting one gratuitous source of confusion, namely, primitive point identities moving around relative to the physical properties.)  There is change if (or where) the metric is not ``stationary,'' that is, if (or where) there exists no time-like Killing vector field  \cite{Wald,ExactSolns}.  (A local section suffices for having no change locally---in a neighborhood where there is a time-like Killing vector, there change does not happen. The Schwarzschild solution outside the horizon is a familiar example.)
A time-like Killing vector field $\xi^{\mu}$ is a vector field such that is Killing relative to the metric, 
\begin{equation} \pounds_{\xi} g_{\mu\nu} = \xi^{\alpha} g_{\mu\nu},_{\alpha} + g_{\mu\alpha} \xi^{\alpha},_{\nu} + g_{\alpha\nu} \xi^{\alpha},_{\mu} = \nabla_{\mu} \xi_{\nu} + \nabla_{\nu} \xi_{\mu} =0, \end{equation} making the metric independent of a coordinate adapted to $\xi^{\alpha}$  \cite[p. 352]{Ohanian},  and that is time-like relative to the metric ($g_{\mu\nu} \xi^{\mu} \xi^{\nu} <0$ using $-+++$ signature). Lie differentiation $\pounds_{\xi}$ is a more or less tensorial directional derivative; when acting on tensors or connections, it gives a tensor \cite{Yano}.
The nonexistence  of  a time-like Killing vector field is the coordinate-invariant statement of change:  change is not being stationary (neighborhood by neighborhood) \cite{ExactSolns}. Quantities are observable if they are tensors, tensor densities, or more general geometric objects \cite{Nijenhuis,Schouten,Anderson} and are not compromised by some other convention-dependence such as electromagnetic gauge freedom.   All of this is  uncontroversial in the 4-dimensional Lagrangian context \cite{Wald}. I mention it in some detail only because of the general failure to attend adequately to the Hamiltonian analogs in the appropriate contexts.

Because change is locally defined in terms of the lack of a time-like Killing vector field (or more generally, the lack of a time-like vector such that the Lie derivative of everything, metrical and material, vanishes, if matter is present), one need not attend to boundary terms to define change.  Thus non-trivial topologies also admit change, even if they have no boundaries.  Even if the world is like a doughnut, there is change in GR. Just watch for change  in a room with no windows, such as a basement. The difficulty in knowing what happens at infinity, or knowing whether there is any such place, will be no hindrance.
 This Lagrangian platitude amounts to a revisionist Hamiltonian research project.


\section{Vacuum GR without Spatial Dependence }

In this section, which forms the heart of the paper, I will show how a proper Hamiltonian treatment of GR does indeed involve change.  Change will be exhibited in a truncated form of GR obtained by simply dropping the spatial dependence.  Such a simplified toy theory has most of the features that threaten to obscure time and change in GR, apart from ``many-fingered time'' (freedom to define various simultaneity hypersurfaces).  The treatment of the toy theory will also show by a tractable and interesting example how  Dirac-Bergmann constrained dynamics  works.  Then I will recall my recent result that a first-class constraint generates a bad physical change---not a gauge transformation as is often held.  Because a first-class constraint does not generate a gauge transformation, one is no longer tempted to infer from the fact that the Hamiltonian of GR is a sum of first-class constraints that time evolution is only a gauge transformation and hence that there is no real evolution or change.  With that erroneous Hamiltonian story undermined, one is well positioned to attend to the correct Hamiltonian story, which involves detailed attention to the equations of motion and to gauge transformations, which must be equivalent to the Lagrangian coordinate transformations (or at any rate to the ones involving time in this spatially truncated toy theory) at least for solutions of the equations of motion.  (Inconveniently, it is provable that kinematically possible but dynamically impossible trajectories have a gauge freedom that is not simply related to changing the time coordinate.)  One already knows from differential geometry what it is for there to be change in a solution of Einstein's equations, that is, for the solution not to be stationary:  change is just the absence of a time-like Killing vector field.  It will be found that the Hamiltonian equations of motion imply that, for all choices of (time) coordinate, something depends on time if and only if there is no (time-like) Killing vector field.  It will then be shown that the gauge generator $G$ implements using Hamiltonian resources exactly the infinitesimal changes of time coordinate \emph{via} Lie differentiation, at least on-shell.


\subsection{Homogeneous Truncation }

One can capture just enough of the real physics of GR, while maintaining manifest relevance and avoiding most complication, by using a toy theory, vacuum GR with all spatial coordinate derivatives dropped. 
Such a restriction is both a substantial physical restriction compared to GR, and an adaptation of coordinates to fit the restricted physics.  Because the coefficient of the shift vector $N^i$ in the Lagrangian, namely the momentum constraint, vanishes when nothing varies with the spatial coordinates, $N^i$  no longer appears in the formalism at all.  Thus no corresponding primary constraints exist.  What remains is reparametrization invariance, with the lapse function $N(t)$ an arbitrary function (assuming positivity, boundedness away from $0,$ differentiability to some order, \emph{etc.} as usual).  Calculations are easy, but the resemblance to GR is clear.  

For a partial treatment of GR in Hamiltonian form, two standard texts treat the subject \cite{MTW,Wald} (but watch for opposite conventions in defining the extrinsic curvature).  These texts take a short-cut by dropping the primary constraints and treating the lapse function and shift vector as Lagrange multipliers with no conjugate momenta---a procedure that is adequate for getting the field equations of these theories, but does not work for general field theories and that leaves one quite unable to express 4-dimensional gauge transformations  in Hamiltonian form.  So one turns elsewhere \cite{Sundermeyer}, especially given the recent $G$-related appreciation of primary constraints.  If one sets $\frac{ \partial}{ \partial x^i}($anything$)=0$ (where i=1,2,3), then the Lagrangian density $\mathcal{L}$---now rather a Lagrangian $L$, simplifies to 
\begin{equation}
L = N \sqrt{h}(K^{ij}K_{ij}-K^2), \end{equation}
where the extrinsic curvature tensor (a sort of 3-dimensional tensorial `velocity' in GR) simplifies due to disappearance of the shift vector and its derivatives:
$$K_{ij} =  \frac{1}{2N} \dot{h}_{ij},$$ with the dot denoting the coordinate time derivative, $h_{ij}$ the spatial 3-metric, and  the lapse function $N$ being related to the space-time metric by $N=\frac{1}{\sqrt{-g^{00} } }.$   It is perhaps better to use the slicing density $\alpha=N/\sqrt{h}$ (or even the logarithm thereof) as a canonical coordinate \cite{YorkHamilton}, partly to enforce positivity \cite{GoldbergKlotz}, but $N$ is quite entrenched. 
The spatial Ricci curvature scalar has disappeared. There being, so to speak, no potential energy, the actual dynamics will be rather dull.  That does not matter, however, because all the formal properties that we need will appear. Note that the time derivative of $N$ does not appear in the Lagrangian.



\subsection{Generalized {Legendre} Transformation from $L$ to $H$}

  Defining canonical momenta as usual, one has
$$\pi^{ij} =_{def} \frac{ \partial L  }{ \partial h_{ij},_0   } =  \sqrt{h}(K^{ij} - K h^{ij})$$ and
$$p=_{def}  \frac{ \partial L  }{ \partial N,_0 }=0.$$ 
In going over to the Hamiltonian formalism, one forgets these definitions, recovering the $\pi^{ij}$ equation as an equation of motion and requiring of $p$ merely that it stay $0$.
That condition is called a ``primary constraint'' because it arises at the first stage of the Hamiltonian analysis.

 Performing the Legendre transformation to get from $L$ to $H$ (and using $p=0$ where needed), one gets the canonical Hamiltonian
\begin{equation} H_c = N \mathcal{H}_0, \end{equation}
where \begin{equation} \mathcal{H}_0   =  \pi^{ij} \pi^{ab} (h_{ia}h_{bj} - \frac{1}{2} h_{ij}h_{ab})/\sqrt{h}. \end{equation}
(Given that the Hamilton density has turned into the Hamiltonian, one could drop the script font for the Hamiltonian constraint $\mathcal{H}_0 $. But to emphasize the link to GR, I will retain it.)    Time evolution is generated by the primary Hamiltonian (which adds the primary constraints to the canonical Hamiltonian \cite{Sundermeyer}:
\begin{equation} H_p = N \mathcal{H}_0 + v p, \end{equation} where $v(t)$ is an arbitrary function, which one readily sees is equated to $\dot{N}$ using the basic Poisson bracket $\{N,p\} =1$ and calculating $\{N,H_p \}.$\footnote{ If one keeps  $\dot{N}$ itself in the formalism \cite{SudarshanMukunda,CastellaniGaugeGenerator}, then one does not suppress $\dot{N}$ \emph{via} the primary constraint $p=0$, introduce a new function $v$ unrelated to phase space, and then recover  $v=\dot{N}.$ }
It is evident that both constraints are first-class.  $\{ \mathcal{H}_0, \mathcal{H}_0 \} =0$ because dropping the spatial dependence makes everything difficult about the `Dirac algebra' disappear and Poisson brackets are anti-symmetric,
in contrast to spatially varying GR's   $\{ \mathcal{H}_0(x), \mathcal{H}_0(y) \} $ which involves derivatives of Dirac $\delta$ functions \cite[p. 239]{Sundermeyer}.  
 $\{p, \mathcal{H}_0 \} =0$ because  $ \mathcal{H}_0$ does not depend on $N.$  Finally,  $\{p,p\}=0$ by antisymmetry and the fact that $p$ is one of the canonical coordinates.   
Letting $\psi$ be an arbitrary function of $h_{ij},$ $\pi^{ij},$ $N,$ and $p$, one can calculate the time evolution of $\psi$:
\begin{eqnarray}
\{ \psi, H_p \} = \frac{ \partial \psi}{ \partial h_{ij} }  \frac{ \partial H_p}{ \partial \pi^{ij} } - \frac{ \partial \psi}{ \partial \pi^{ij} } \frac{ \partial H_p}{ \partial h_{ij} } +  \frac{ \partial \psi}{ \partial N} v   - \frac{ \partial \psi}{ \partial p}  \cdot 0 
\end{eqnarray} using  $ \mathcal{H}_0 = 0$ after taking the Poisson bracket.  (The symbol $\approx$ is often used for equality on a constraint surface, but I will not bother. One also speaks of equations that hold ``on-shell,'' which is shorter than ``on the constraint surface and using the evolution equations as needed.'')  
It follows that \begin{eqnarray}
\{ h_{ij}, H_p \} =  \frac{ \partial H_p}{ \partial \pi^{ij} }   = \frac{2N}{\sqrt{h} }(h_{ai}h_{bj} - \frac{1}{2} h_{ij}h_{ab}) \pi^{ab} = \dot{h}_{ij},  \nonumber \\
\{ \pi^{ij}, H_p \} = - \frac{ \partial H_p}{ \partial h_{ij} }  =  -\frac{N}{\sqrt{h} }(2\pi^{ia}\pi_{a}^{j}-\pi \pi^{ij} -\frac{1}{2}\pi^{ab}\pi_{ab}h^{ij} +\frac{1}{4} \pi^2 h^{ij})  = \dot{\pi}^{ij},  \nonumber \\
\{ N, H_p \} = \frac{\partial H_p}{\partial p}= v=\dot{N},  \nonumber \\
\{p, H_p \} =  -\frac{ \partial H_p}{ \partial N }   = -\mathcal{H}_0 =0= \dot{p}.  
\end{eqnarray}


\subsection{A First-Class Constraint Generates a Bad Physical Change}

Much as in the electromagnetic case, in GR or its homogeneous reduction a first-class constraint generates not a gauge transformation, but a bad physical change.  (One expects such a result to be typical for theories in which primary first-class constraints have descendents, secondary and maybe higher generations of constraints that follow from the first-class primaries.)  
In the case of electromagnetism \cite{FirstClassNotGaugeEM} and full GR (in preparation \cite{FirstClassNotGaugeGR}), one way to spot the bad physical change is in the spoilage of the Lagrangian constraint equations, Gauss's law in terms of the electric field (not canonical momenta) or the Gauss-Codazzi relations describing the embedding of space into space-time. Such spoilage of a Lagrangian constraint is a sufficient condition for a change to be bad, but presumably is not necessary.

What does $p$ do to the $q-\dot{q}$ Hamiltonian constraint, the normal-normal part of Einstein's equations?
To answer that question in GR, it is convenient to define a lapse-less factor in the extrinsic curvature tensor:
$L_{ij} =_{df} N K_{ij} = \frac{1}{2}(\dot{h}_{ij} -D_i N_j -D_j N_i).$  
 Thus  \begin{eqnarray}  \{  \int d^3y \epsilon(y) p(y), K^{ij} K_{ij} -K^i_i K^j_j - R(x) \} = \int d^3y \epsilon(y) \{ p(y), N^{-2}(x) (L^{ij} L_{ij} -L^2(x)) \} \nonumber \\  = 2 \epsilon(x) N^{-1}(K^{ij} K_{ij} -K^2), \label{Violate} \end{eqnarray}
which in full GR is generally nonzero due to the spatial Ricci scalar term $-R.$ 
In the homogeneous truncation,  $R$ and the shift vector $N^i$ are absent.  But the former contributes nothing anyway, while the latter's contribution (by virtue of being multiplied by a factor involving the lapse) is accounted for in the $K_{ij}$ terms, so the absence of those contributions in the homogeneous truncation does not change the result:
 \begin{eqnarray}  \{  \int d^3y \epsilon(y) p(y), K^{ij} K_{ij} -K^i_i K^j_j \}  = 2 \epsilon(x) N^{-1}(K^{ij} K_{ij} -K^2). \end{eqnarray} 
In GR, that expression would be generically nonzero (and not a Lie derivative either) because $K^{ij} K_{ij} -K^i_i K^j_j = R$; hence that Poisson bracket would clearly manifest a bad physical change, spoiling the constraint equation that is the time-time part of  Einstein's equations.  
For the homogeneous truncation, the vanishing of the 3-dimensional Ricci scalar $R$ makes that Poisson bracket look not so bad.  Sometimes very special cases produce potentially misleading simplifications, so it is best not to forget the guidance from full GR, given that it (not the homogeneous truncation) is ultimately what we seek to understand.  For GR, one can show that each of the first-class constraints $p,$ $p_i,$ $\mathcal{H}_0$ and $\mathcal{H}_i$ spoils all of the Lagrangian constraints. 
 One expects that the appropriate teaming arrangements (the gauge generators) will cause all the badness to cancel out, which has been show to occur in one case and presumably will occur when the remaining three are completed.
 The calculations are analogous to the electromagnetic case except, of course, much harder.  
Much as one notices for electromagnetism that the electric field is changed by the primary first-class constraint and \emph{therefore} Gauss's law is violated, spoilage of the Lagrangian constraints is a generally logically weaker consequence of first-class constraint transformation behavior that differs from the known gauge transformations.

One can go on to show that varying the lapse function $N$ in an arbitrary way changes what ought to be physically invariant quantities, namely, Komar's intrinsic Weyl curvature scalar coordinates.  (Here one assumes a generic but not completely general situation, namely, the absence of Killing vector fields; see also \cite{KomarScalars}.)  The Weyl tensor, the totally traceless part of the Riemann curvature tensor for space-time, is the part of the Riemann tensor not specified by Einstein's equations.  Typically it has 4 independent scalar concomitants, two quadratic, two cubic \cite{BergmannKomarPRL,BergmannHandbuch,PonsSalisbury}, that remain after Einstein's equations (source-free, coupled to electromagnetism, or perhaps more generally) have been imposed.  One can of course devise other functions of these \cite{BergmannKomarPRL}.  (In that sense, any coordinate system is a Weyl scalar coordinate system, a fact that has significant implications.)  The Weyl scalars can be written, using Einstein's equations if necessary, in terms of the spatial metric tensor, the extrinsic curvature tensor, and their derivatives.  In terms of canonical coordinates, one can re-express these expressions in terms of the  canonical momenta $\pi^{ij}$ instead of the extrinsic curvature $K_{ij}.$ In terms of the 3-metric and its canonical momenta, the Weyl scalars are independent of the lapse and the shift vector.   But such replacement uses the Hamilton equations $\frac{\delta H_p}{\delta \pi^{ij} } = \dot{h}_{ij},$ which are not identities.  Hence a more primordial form of the Weyl scalars leaves them in terms of $h_{ij}$ and $K_{ij}.$  But $K_{ij}$ is not primitive; it is derived in terms of the definition  $K_{ij} =  \frac{1}{2N} \dot{h}_{ij}+\ldots.$  Hence the Weyl scalars in terms of $q$ and $\dot{q}$ depend on $N$ inversely from how they depend on  
the lapse-independent quantity $L_{ij} = N K_{ij},$ which depends on $\dot{h}_{ij}.$  Hence one could write the Weyl scalars in terms of $h_{ij},$ $L_{ij},$ and $N$ in order to make the dependence on $q$ and $\dot{q}$ clearer.  But then it is evident that varying $N$ arbitrarily (while leaving $\dot{h}_{ij}$ alone), as the smeared primary constraint $p$ does, will \emph{alter} the Weyl scalar(s)---something that a real gauge transformation could never do.  Hence what the primary first-class constraint $p$ generates is a bad physical change, not a gauge transformation.

With that reminder to check the full transformation behavior, not just what happens to the Lagrangian constraints, one can consider the Hamiltonian constraint  $\mathcal{H}_0$.    One can find the relation between $\mathcal{H}_0$ and space-time coordinate transformations by starting with the gauge generator $G$ \cite{CastellaniGaugeGenerator} and throwing away some terms to isolate $\mathcal{H}_0$.  As will appear below, $G$ has a bunch of terms involving the primary constraints, the lapse and shift, and (in some cases) the spatial 3-metric \cite{CastellaniGaugeGenerator}; these will not affect the 3-metric $h_{ij},$ which only sees it own conjugate momentum due to the structure of the Poisson bracket. Homogeneous truncation also eliminates $\mathcal{H}_i.$ Thus $$ \{ h_{ij}(x), G \} = \{ h_{ij}(x), \int d^3y  \epsilon^{\perp}(y) \mathcal{H}_0(y)  \},$$ where $\epsilon^{\perp}$ is the normal projection of the 4-vector $\xi^{\mu}$ describing an infinitesimal coordinate transformation,  $\epsilon^{\perp} = N \xi^0$.  
One has 
$$  \{ h_{ij}(x), \int d^3y \epsilon^{\perp}(y) \mathcal{H}_0(y) \} =    \delta^{\mu}_i \delta^{\nu}_j \pounds_{(\epsilon^{\perp} n^{\alpha}) } g_{\mu\nu}(x). $$ That looks good for getting a coordinate transformation out of $\mathcal{H}_0$, but then $\{ N(x), \int d^3y \epsilon^{\perp}(y) \mathcal{H}_0(y) \}=  0$ shows that part of the remainder  of the space-time Lie derivative formula is violated:  the lapse $N$ certainly changes under a change of time coordinate, but $\mathcal{H}_0$ fails to effect such a change.  Thus $ \mathcal{H}_0$ does not generate a coordinate transformation either.  As with electromagnetism, a first-class constraint, primary or secondary, generates a bad physical change.  Only the special combination $G$ generates a gauge transformation.


%

\subsection{Quasi-Invariance of Action \emph{vs.} First-Class Constraints}

Above it was found for electromagnetism that each first-class constraint changes the action, but the same specially tuned sum that preserves the electric field also preserves the action (up to a boundary term), yielding agreement between two different criteria.  For GR it is less clear what counts as an observable, the analogous argument to preserving the electric field might be problematic.  One can still test each individual first-class constraint for whether it preserves the action (up to a boundary term), however.  As with electromagnetism, neither one does.  One can also see that a specially tuned sum, the gauge generator $G$, does change the action  by only a boundary term.

The canonical Lagrangian  $ L_H = p \dot{q}-H = p \dot{N} + \pi^{ij} \dot{h}_{ij} - H_{p} = \pi^{ij} \dot{h}_{ij} - H_{canonical} =  \pi^{ij} \dot{h}_{ij} - N \mathcal{H}_0$ introduces canonical momenta $\pi^{ij}$ as auxiliary fields into the action, thus preserving equivalence to the usual Lagrangian  $ L.$  To see what the primary constraint $p$ does, one multiplies it by an arbitrary function $\xi(t)$ (though without any spatial smearing):
\begin{eqnarray}
\{ \xi p, \pi^{ij} \dot{h}_{ij} - N \mathcal{H}_0 \} = 0 + \{ \xi p,  - N \mathcal{H}_0 \} = \xi \mathcal{H}_0,\end{eqnarray}
which is not a total derivative.  Thus the primary first-class constraint does not generate a gauge transformation.

Likewise for the secondary constraint:
\begin{eqnarray} \{ \epsilon(t) \mathcal{H}_0, \pi^{ij} \dot{h}_{ij} - N \mathcal{H}_0 \} = \{  \epsilon \mathcal{H}_0, \pi^{ij} \dot{h}_{ij} \} - 0 =
\{  \epsilon \mathcal{H}_0, \pi^{ij} \}  \dot{h}_{ij} + \{  \epsilon \mathcal{H}_0, \dot{h}_{ij} \} \pi^{ij}.
\end{eqnarray}
One can evaluate this expression either by integrating over time and integrating by parts, or using the Anderson-Bergmann velocity Poisson bracket.  Either way, one finds the result 
\begin{eqnarray} \{ \epsilon(t) \mathcal{H}_0, \pi^{ij} \dot{h}_{ij} - N \mathcal{H}_0 \} = \epsilon \dot{\mathcal{H}}_0 - \frac{\partial}{\partial t} \left(\epsilon \pi^{ij} \frac{ \partial \mathcal{H}_0 }{ \partial \pi^{ij} } \right), \end{eqnarray}
which is not a total derivative.  Thus the secondary  first-class constraint does not generate a gauge transformation either.

If one combines the primary and secondary constraints with their arbitrary coefficients, and then asks the sum to be a total derivative, one can succeed; one finds that $\xi=\dot{\epsilon}$ and the canonical Lagrangian $p \dot{q}-H$ changes by $$\frac{\partial}{\partial t} \left(\epsilon \left[\mathcal{H}_0 - \pi^{ij}  \frac{\partial \mathcal{H}_0 }{\partial \pi^{ij} } \right] \right).$$ Thus the action is quasi-invariant, which is just the condition for preserving the equations of motion.   The tuned sum of first-class constraints is, not surprisingly, just the gauge generator $G$.  

%
%


\subsection{Change from Hamilton's Equations}  %

Change is (real) time dependence. There is a risk of fake change in GR by a funny time choice  or a funny labeling of space over time.  Hence one needs a savvy definition:  there is real change if and only if, for \emph{all} choices of coordinates, there is time dependence.  In reparametrization-invariant theories, one can try to generate fake change by speeding up (bunching up) or slowing down (spreading out) the labeling of the time slices. In GR the possibilities for fake change are much more varied due to many-fingered time and to the possibility of letting spatial coordinates slide around over time \emph{via} the shift vector.  But I'll continue with the toy theory that is simple, but not too simple. Fake change is then apparent change that exists only for some choices of time coordinate. Note that the question is whether \emph{something or other} depends on time; it isn't obviously required that some single quantity depend on time in all coordinate systems.


  If there is real change, then there is time dependence for all choices of labeling of the time slices in that interval: 
$ \forall $ labelings $(\dot{h}_{ij} \neq 0  \hspace{.1in}  \lor \hspace{.1in} \dot{\pi}^{ij}  \neq 0  \hspace{.1in} \lor  \hspace{.1in} \dot{N} \neq 0).$  
(If there is no real change, then 
$\exists$ labeling such that $\dot{h}_{ij} =0 \hspace{.1in} \& \hspace{.1in} \dot{\pi}^{ij} =0 \hspace{.1in} \& \hspace{.1in} \dot{N}=0.$)  
It would be nice if one didn't have to worry about $\dot{N},$ a rather slippery quantity that is determined by conventional gauge (coordinate) choice: it vanishes for some time coordinates but not others.  Thus $\dot{N} \neq 0$ doesn't hold reliably; but can $\dot{N} $  team up with $\dot{h}_{ij}$ or $ \dot{\pi}^{ij} $ so that at least one of them is nonzero for any time coordinate?   Not usefully:   $\dot{h}_{ij},$  $\pi^{ij}$, and $\dot{\pi}^{ij}$ are all scalar densities (of whatever weights) or scalars  under change of time coordinate.  Thus  if they vanish, they do so invariantly; if they fail to vanish, they do so invariantly. So they can't use help from $\dot{N}.$ 
If there is a time coordinate such that  $\dot{h}_{ij} =0$ and $  \dot{\pi}^{ij}  = 0,$ there is a time coordinate that also yields   $\dot{N}=0$ because $\dot{N}$ has an affine rather than strictly linear time coordinate transformation law.  It therefore cannot bear the load of being the locus of real change, and indeed cannot even help to bear that load. 
So the quantifier over time coordinates (labelings of the slices) can be dropped, as can the disjunct $\dot{N} \neq 0.$
 Thus there is a nice
\begin{quote} {\bf Result:}  There is   real change in the homogeneous vacuum theory  if and only if $\dot{h}_{ij} \neq 0 \hspace{.1in} \lor \hspace{.1in} \dot{\pi}^{ij}  \neq 0$ in some  coordinate system (and hence in all coordinate systems). 
\end{quote} 
One could simplify a bit further using Hamilton's equations for the toy theory, but there is no need to do so and some of the simplifications would not carry over to GR.

  The reasoning also is reversible.  Because in the homogeneous truncation $h_{ij}$ and $\pi^{ij}$ are a scalar and a pseudo-scalar (respectively) under changes of time coordinate (hence scalars for the infinitesimal transformations used in Lie differentiation), the non-vanishing of either of these velocities  $\dot{h}_{ij} \neq 0 $ or   $ \dot{\pi}^{ij}  \neq 0$  in some coordinate system implies that same non-vanishing in all time coordinate systems, and thus the non-vanishing of the time derivative of something or other in every coordinate system.

%


\subsection{Change from Differential Geometry:  No Time-like Killing Vector}

Now let us ascertain the conditions for the non-existence or existence of a time-like Killing vector, or the relevant analog thereof, and see how it lines up with the time evolution generated by $H_p$.
We need the $3+1$ split of the Lie derivative formula for an infinitesimal coordinate transformation
 $\pounds_{\xi} g_{\mu\nu} = \xi^{\alpha} g_{\mu\nu},_{\alpha} + g_{\mu\alpha} \xi^{\alpha},_{\nu} + g_{\alpha\mu} \xi^{\alpha},_{\mu}$ \cite{MukhanovWipf}.
Actually we need only the $x^i$-independent homogeneous truncation of the $3+1$ split, which is much cleaner.   
One has, besides rigid affine spatial coordinate transformations (which would be important if we cared about counting degrees of freedom \cite{RyanShepley,AshtekarBianchi,PonsReduce}, but which I ignore),
 $$ \delta N = \xi^0 \dot{N} + N \dot{\xi^{0}}, $$  $$ \delta h_{ij} = \xi^0 \dot{h}_{ij}.$$
The formula for the variation of the momenta 
$\delta \pi^{ij} $ is derived using $K_{ij} =N \Gamma^0_{ij}$ and Hamilton's equations  \cite{PonsSalisburyShepleyYang,ThiemannBook}. The result simplifies enormously for the homogeneous toy theory: $\delta \pi^{ij} = \xi^0 \dot{\pi}^{ij} $.    One also needs $p$ to stay $0$ somehow. 
One can show that $p$ is a weight $-1$ scalar density under change of time coordinate, so  $\delta p = \pounds_{\xi} p =  \xi^0 \dot{p} - \dot{\xi}^0 p,$ which will stay $0$ using both the constraints.   

Hence the condition for the existence of a time-like Killing vector field is equivalent (using Hamilton's equations as needed) to
\begin{eqnarray} (\exists \xi^0) (\xi^0 \dot{N} + N \dot{\xi^{0}} =0 \hspace{.5in} \& \hspace{.5in}  \xi^0 \dot{h}_{ij} = 0 \hspace{.5in} \& \hspace{.5in} 
 \xi^0 \dot{\pi}^{ij}=0). \end{eqnarray}  Thus the condition for change is just the negation of this triple conjunction. Below some logical manipulations will be performed from this point.


\subsection{$G$ Generates  Lie Differentiation On-shell in Hamiltonian Formalism}

The gauge generator $G$ by design  ought to give these same formulas just obtained from 4-dimensional differential geometry, at least for solutions of all the Hamiltonian equations and constraints.  
$G$ is a bit complicated in GR  \cite{CastellaniGaugeGenerator,PonsSalisburyShepleyYang}:
it is a specific sum of the secondary constraints $\mathcal{H}_0$ and $\mathcal{H}_i$ \emph{and} the primary constraints $p$ and $p_i$ with coefficients savoring of Lie differentiation. While it will not be necessary here to use the whole expression, it might be useful to have it anyway. Making sure that $G$ and the changes that it generates live in phase space requires eliminating $\dot{N}$ by taking not the time component $\xi^0,$ but a $3+1$ projected relative $\epsilon^{\perp} = N \xi^0 $ as primitive and hence having vanishing Poisson brackets with everything \cite{CastellaniGaugeGenerator,PonsSalisburyShepleyYang}. 
(My notation, opposite to Pons, Salisbury and Shepley's but partly in line with Castellani's, uses $\xi^{\mu}$ for the coordinate basis components  of the vector $\xi$ and lets $\epsilon$ be the $3+1$ projected descriptors.)

 It is most readily given in two parts, one for gauge transformations that do not preserve the simultaneity hypersurface, and one for spatial coordinate transformations. Note that I am not worrying about spatial boundary terms (on which the two sources above disagree), partly because there is no point in trying to run before walking.  Castellani's notation also tends to hide dependence on the spatial metric, which is made explicit here.   The normal gauge generator is
\begin{eqnarray}
G[\epsilon, \dot{\epsilon}] = \int d^3x [\epsilon^{\perp} \mathcal{H}_0  + \epsilon^{\perp} p_j h^{ij} N,_i + \epsilon^{\perp}(N p_i h^{ij}),_j + \epsilon^{\perp} (p N^j),_j + \dot{\epsilon}^{\perp} p]. \label{normalgaugegenerator}
\end{eqnarray}
It generates on phase space-time a transformation that, for solutions of the Hamiltonian field equations, changes the time coordinate in line with 4-dimensional tensor calculus. 
The spatial gauge generator is
\begin{eqnarray}
G[\epsilon^i, \dot{\epsilon}^i] = \int d^3x [\epsilon^i \mathcal{H}_i  +  \epsilon^i N^j,_i p_j - \epsilon^j,_i N^i p_j  + \epsilon^i N,_i p + \epsilon^i,_0 p_i]. \label{spatialgaugegenerator}
\end{eqnarray}
This spatially projected descriptor $\epsilon^i$, on account of the shift vector that allows the spatial coordinates to slide around over time, is not simply the spatial components $\xi^i$ of $\xi$.  Instead it is given by 
$ \epsilon^i = \xi^i+ N^i \xi^0.$ It generates spatial coordinate transformations that respect  the simultaneity hypersurfaces.

Roughly speaking, the coefficients of the primary constraints are those needed to transform the lapse and shift in such a way as to fill the holes left by the secondary constraints (\emph{c.f.} the common error that the secondary constraints by themselves do generate coordinate transformations).  For spatial coordinate transformations, which have the virtues of being defined even without using the equations of motion and manifestly relating to fragments of tensor calculus, one has the good news \cite[p. 241]{Sundermeyer}
\begin{eqnarray}
\{ h_{ij}(x), \int d^3y \xi^k(y) \mathcal{H}_k(y) \} = \pounds_{\xi} h_{ij}(x), \nonumber \\
\{ \pi^{ij}(x), \int d^3y \xi^k(y) \mathcal{H}_k(y) \} = \pounds_{\xi} \pi^{ij}(x) 
\end{eqnarray}
that might tempt one to think that a coordinate transformation is being made, but also the bad news 
\begin{eqnarray} 
\{ \mathcal{H}_i(x), N^j(y) \}=0, \nonumber \\
\{ \mathcal{H}_i(x), N(y) \}=0
\end{eqnarray}
that shows that no coordinate transformation is made unless quite specific assistance from the primary constraints $p_i$ and even $p$  is brought in.  A similar story holds for $\mathcal{H}_0$ \cite[p. 603]{KiriushchevaMyths}, which is loosely tied to both time evolution and changes of time coordinate, but by itself generates neither one.  

 In our $x^i$-independent homogeneous truncation, $G$  simplifies nicely. One can ignore the spatial gauge generator.   Throwing away spatial dependence, besides discarding the momentum constraint $\mathcal{H}_i=0$ and the primary constraints $p_i=0$ tying down the momenta conjugate to the shift vector, also annihilates the
 structure constant $C^0_{00}=0$ for the Hamiltonian constraint with itself. 
 One has for the spatially truncated normal gauge generator  
\begin{equation}  
G =  \epsilon^{\perp} \mathcal{H}_0 + p\dot{ \epsilon}^{\perp}. 
\end{equation} 
One finds that $G$ has the following Poisson brackets with the basic canonical variables: 
  \begin{eqnarray}
\{ h_{ij}, G \} =  \xi^0 N \frac{2}{\sqrt{h} }(h_{ai}h_{bj} -\frac{1}{2} h_{ij}h_{ab}) \pi^{ab} = \xi^0 \{ h_{ij}, H_p \},           \nonumber \\
\{ \pi^{ij}, G \} =     -  \xi^0   \frac{N}{\sqrt{h} }(2\pi^{ia}\pi_{a}^{j}-\pi \pi^{ij} -\frac{1}{2}\pi^{ab}\pi_{ab}h^{ij} +\frac{1}{4} \pi^2 h^{ij})= \xi^0 \{ \pi^{ij}, H_p \},  \nonumber \\
\{ N, G \} =  \dot{\epsilon^{\perp} } = \xi^0 \dot{N} + \dot{\xi^0 } N,    \nonumber \\ 
\{p, G \} =   0  
\end{eqnarray}
identically.
 Using Hamilton's equations, one has ``on-shell'' 
\begin{eqnarray}
\{ h_{ij}, G \} =  \xi^0 \dot{h}_{ij},  \nonumber \\ 
\{ \pi^{ij}, G \} = \xi^0 \dot{\pi}^{ij},  \nonumber \\
\{ N, G \} = \xi^0 \dot{N} + \dot{\xi^0 } N, \nonumber \\ 
\{p, G \} = 0.   
\end{eqnarray}
One sees that these equations match the Lie derivative formulas $ \delta N = \xi^0 \dot{N} + N \dot{\xi^{0}}$ ($N$ being a weight $1$ scalar density under change of time coordinate and thus having this Lie derivative \cite{Anderson}), $\delta h_{ij} = \xi^0 \dot{h}_{ij},$  and $\delta \pi^{ij} = \xi^0 \dot{\pi}^{ij} $,  while $\delta p = 0$ is also fine even without fitting the weight $-1$ density character of $p$, so the gauge generator $G$ deserves its name.  The Hamiltonian formalism  implements 4-dimensional coordinate transformations at least on solutions of the Hamilton equations 
 \cite{ThiemannBook}; here the one-dimensional temporal analog is explicit and convenient.

If and only if there exists a (time-like) Killing vector `field' $\xi^{\mu}$, 
the Poisson brackets of the canonical variables with $G$ should all vanish (in all time coordinates):
\begin{eqnarray}
\{ h_{ij}, G \} =  \xi^0 \dot{h}_{ij} =0,  \nonumber \\ 
\{ \pi^{ij}, G \} = \xi^0 \dot{\pi}^{ij} =0,  \nonumber \\
\{ N, G \} = \xi^0 \dot{N} + \dot{\xi^0 } N=0, \nonumber \\ 
\{p, G \} = 0.   
\end{eqnarray}
This is just the Killing vector condition \begin{eqnarray} (\exists \xi^0) (\xi^0 \dot{N} + N \dot{\xi^{0}} =0 \hspace{.1in} \& \hspace{.1in} \xi^0 \dot{h}_{ij} = 0 \hspace{.1in} \& \hspace{.1in}
 \xi^0 \dot{\pi}^{ij}=0) \end{eqnarray}  from the previous subsection, along with a suitable claim about the primary constraint (the boring canonical momentum that is always $0$).
Change is related to a \emph{lack} of a time-like Killing vector field, so let us negate. The primary constraint $p$ has to remain $0$ no matter what.  The Killing vector condition, being tensorial, holds in all coordinate systems or fails in all of them, so there is no need to quantify over labelings (time coordinates).  Thus  the lack of a Killing vector is $$(\forall \xi^0) (\xi^0 \dot{h}_{ij} \neq 0  \hspace{.1in} \lor \hspace{.1in} \xi^0 \dot{\pi}^{ij}  \neq 0 \hspace{.1in} \lor  \hspace{.1in} \xi^0 \dot{N} + \dot{\xi^0 } N \neq 0).$$ 
Vanishing $ \xi^0$ would not count as Killing (or time-like), so the no-Killing condition is 
 $$(\forall \xi^0) (\dot{h}_{ij} \neq 0  \hspace{.1in} \lor \hspace{.1in}  \dot{\pi}^{ij}  \neq 0 \hspace{.1in} \lor  \hspace{.1in} (\xi^0 N),_0 \neq 0).$$ 
All three disjuncts are scalar densities of some weight or other, so their vanishings or not are invariant---recall that no quantification over time labelings was needed.  
There is always some $\xi^0$ that can make  $(\xi^0 N),_0 = 0$; $\xi^0 = N^{-1}$ or some constant multiple thereof will do. Because the last disjunct $(\xi^0 N),_0 \neq 0$ is unreliable and the other two don't depend on $\xi^0$, the quantification over $\xi^0$ can also be dropped.  
Hence one has another 
\begin{quote} {\bf Result:}  the non-existence of a time-like Killing vector field is equivalent to  $\dot{h}_{ij} \neq 0 \hspace{.07in} \lor \hspace{.07in} \dot{\pi}^{ij} \neq 0.$ \end{quote}
 That condition holds in all (time) coordinates if it holds in any.  This reasoning is also reversible.

This condition is exactly the one found above using the time evolution generated by $H_p$ and asking it to be nonzero in all coordinate systems. Thus the primary Hamiltonian's time evolution gives exactly the same result as 4-dimensional differential geometric Lie differentiation. The primary Hamiltonian is thus vindicated out of the right choice out of the six or more candidates---no surprise in light of the known equivalence of the primary Hamiltonian to the Lagrangian.

 The Lie derivative formula, in turn, is implemented in the Hamiltonian formalism (at least on-shell) using the gauge generator $G$, as was already known \cite{CastellaniGaugeGenerator}. There is real change just in case there is no time-like Killing vector field.  Expecting the Hamiltonian formalism to match the unproblematic Lagrangian/differential geometric formalism has resolved the problem in terms of the role of a time-like Killing vector field, as promised in the title.
This agreement makes change in classical canonical GR, or at any rate in the toy theory, luminously clear and satisfying.  
Working in full GR would add messy terms that tend to obscure the point.  GR adds the further issue of many-fingered time.  I expect that an analysis of this sort would work fine even for GR.

Defining change in terms of the lack of a time-like Killing vector field provides an attractive way to remain non-committal regarding a choice between, for example, ``intrinsic time'' involving $h$ or ``extrinsic time'' involving $\pi^{ij}  h_{ij},$ at least classically.   The expressions involving quantifiers and conjunctions ($\&,$ and) or disjunctions ($\lor$, or) give a tensorial statement that $h_{ij}$ or $\pi^{ij}$ (or both) depends on time $t.$  It might well be the case that a time could be dug out of $h_{ij}$ in some regions but not others, but that $\pi^{ij}$ can fill the gaps; a recollapsing Big Bang model at the moment the expansion stops and reverses is a familiar example. Of course it would be ideal not to have to choose at all.  But if one must choose, the equivalence of the expressions using quantifiers and conjunctions or disjunctions to the (negation of the) tensorial Killing equation implies that there is always something that could be chosen as time, and that the baton is smoothly passed back and forth as needed (classically).


\section{Change with Matter:  A Massive Scalar Field}


\subsection{Change in Hamiltonian Formalism from $H_p$ }

Thus far change has been sorted out for (a homogeneous truncation of) vacuum General Relativity, but not for (a homogeneous truncation of) General Relativity with sources.  To address
the latter question, one can introduce a scalar field $\phi$, for generality a massive scalar field.  Recalling that change is ineliminable time dependence, there is no change in a space-time region  just in case there exists everywhere in that region  a time-like vector field $\xi^{\mu}$ that is `Killing' in a generalized sense for both the metric and the matter:
\begin{eqnarray}
 (\exists \xi^{\mu}) (\pounds_{\xi} g_{\mu\nu}=0 \hspace{.1in} \&  \hspace{.1in}    \pounds_{\xi} \phi =0).
\end{eqnarray}
What does this condition come to in terms of a $3+1$ Hamiltonian formalism?
It will give a Hamiltonian definition of change, one potentially involving the matter as a change-bearer, not just gravity.  As usual, throwing away spatial dependence will simplify matters.

The starting Lagrangian density is now \cite{Sundermeyer}
\begin{equation} 
\mathcal{L} = \mathcal{L}_{GR} -\frac{1}{2} \phi,_{\mu} \phi,_{\nu} g^{\mu\nu} \sqrt{-g} -\frac{m^2}{2} \phi^2 \sqrt{-g}.
\end{equation}
Making the 3+1 ADM split and discarding spatial dependence, one has
\begin{eqnarray}
L=N \sqrt{h}(K^{ij}K_{ij} -K^2) + \frac{ \sqrt{h} }{2N} \dot{\phi}^2 -\frac{ m^2}{2} N \sqrt{h} m^2 \phi^2.
\end{eqnarray}
The canonical momentum $\pi^{ij}$ for gravity is as before, while the new canonical momentum for the scalar field $\phi$ is
\begin{equation}  \pi_{\phi}= \frac{ \partial L}{\partial \dot{\phi} } = \frac{ \sqrt{h} }{N} \dot{\phi },
\end{equation}
which is trivially inverted. 
The primary Hamiltonian is
\begin{eqnarray}
H_p = N(\mathcal{H}_0 + \mathcal{H}_{0\phi}) + v p =  N\mathcal{H}_0 + N\left( \frac{ \pi_{\phi}^2 }{2 \sqrt{h} } + \frac{ m^2 \sqrt{h} \phi^2 }{2}\right) + v p.
\end{eqnarray}

Note that a massless scalar field ($m=0$) would behave as a free particle, making $\phi$ evolve monotonically and hence be a pretty good clock, but a massive scalar field behaves as a harmonic oscillator, with the $\phi$ and its momentum $ \pi_{\phi}$ oscillating.  Hence a massive scalar field avoids unrealistic simplicity and thus is more representative of other matter fields and even what happens in spatially inhomogeneous contexts than is a massless scalar field.

One can now find Hamilton's equations.
$$ \{ h_{ij}, H_p \} =  \frac{ \partial H_p}{ \partial \pi^{ij} }   = \frac{2N}{\sqrt{h} }(h_{ai}h_{bj} - \frac{1}{2} h_{ij}h_{ab}) \pi^{ab} = \dot{h}_{ij}$$ as before.
$$ \{ \pi^{ij}, H_p \} = - \frac{ \partial H_p}{ \partial h_{ij} }  =  -\frac{N}{\sqrt{h} }(2\pi^{ia}\pi_{a}^{j}-\pi \pi^{ij} -\frac{1}{2}\pi^{ab}\pi_{ab}h^{ij} +\frac{1}{4} \pi^2 h^{ij}) + \frac{N}{4\sqrt{h}}\left(\pi_{\phi}^2 h^{ij} - m^2 h h^{ij} \phi^2  \right)  
  = \dot{\pi}^{ij}. $$
$ \{ N, H_p \} = \frac{\partial H_p}{\partial p}= v=\dot{N}$ as before. 
$ \{p, H_p \} =  -\frac{ \partial H_p}{ \partial N }   = -\mathcal{H}_0 - \mathcal{H}_{0\phi} = 0 = \dot{p}, $ sprouting a contribution from $\phi$.
The novel equations are $  \{ \phi, H_p \} = \frac{ N \pi_{\phi}  }{ \sqrt{h} } = \dot{\phi}, $ which inverts the Legendre transformation back from $q$, $p$  to $q,$ $\dot{q}$ for matter $\phi$,  and
\begin{eqnarray}  \{ \pi_{\phi}, H_p \}  = -m^2 N \sqrt{h} \phi = \dot{\pi}_{\phi},  \end{eqnarray}
which gives the interesting part of the dynamics of the massive scalar field.

There is real change if and only if something depends on time for every choice of time coordinate (labeling):
$$(\forall labeling)(\dot{h}_{ij} \neq 0 \hspace{.1in} \lor \hspace{.1in} \dot{\pi}^{ij} \neq 0 \hspace{.1in} \lor \hspace{.1in} \dot{N}\neq 0 \hspace{.1in} \lor \hspace{.1in} \dot{\phi} \neq 0 \hspace{.1in} \lor \hspace{.1in} \dot{\pi}_{\phi} \neq 0).$$
One can always set $\dot{N}$ to $0$, because its transformation law is affine, but 
$\dot{h}_{ij},$ $\dot{\pi}^{ij},$ $\dot{\phi},$ and $\dot{\pi}_{\phi}$ are all scalar densities, vanishing or not invariantly.  Hence one can drop both the $\dot{N}$ disjunct and the quantifier $\forall$ over time labelings.
Thus there is change iff 
\begin{eqnarray} \dot{h}_{ij} \neq 0 \hspace{.1in} \lor \hspace{.1in} \dot{\pi}^{ij} \neq 0 \hspace{.1in} \lor \hspace{.1in} \dot{\phi} \neq 0 \hspace{.1in} \lor \hspace{.1in} \dot{\pi}_{\phi} \neq 0.
\end{eqnarray}
Thus the burden of bearing change can be shared among these four quantities.  One can, for example, pass the baton around as needed among  $h_{ij},$ $\pi^{ij},$  $\phi$ and $\pi_{\phi}$.


\subsection{Differential Geometric Change: No Generalized Killing Vector}

From the standpoint of differential geometry, a solution of Einstein's equations with a scalar field should be regarded as changeless (a generalization of stationarity) just in case there is no time-like vector field $\xi^{\mu}$ satisfying the generalized Killing condition
\begin{eqnarray}  
\pounds_{\xi} g_{\mu\nu}=0  \hspace{.1in} \&  \hspace{.1in}
\pounds_{\xi} \phi=0.
\end{eqnarray}
Neither gravity nor matter changes, and nothing else is present, so nothing changes.  
In principle this pair of equations might be redundant if the equations are not independent. Can matter change without making gravity change also?  That is of no concern for present purposes, partly because such a result might be model-dependent:   it might depend on what types and numbers of fields are used as matter.  What is of interest is not special features of a massive scalar field, but features likely to be representative of a broad class of 
matter sources in GR.

As before, a $3+1$ split, followed by throwing away spatial dependence, is useful.  The nontrivial new part is that we will need something like 
`$\pounds_{\xi} \pi_{\phi}$.'  From the experience with the vacuum case, we expect to need the 
relation $ \pi_{\phi}=  \frac{ \sqrt{h} }{N} \dot{\phi },$    which holds due to 
 $$  \{ \phi, H_p \} = \frac{ N \pi_{\phi}  }{ \sqrt{h} } = \dot{\phi}. $$
Under time relabelings, $\sqrt{h}$ is unmoved, whereas $N^{\prime} = N \left| \frac{ \partial x^0}{ \partial x^{0\prime } } \right| $ (weight $1$ and not flipping signs under time reversal)  and  $\dot{\phi}^{\prime} = \dot{\phi}  \frac{ \partial x^0}{ \partial x^{0\prime } } $ (also weight $1$ but flipping signs under time-reversal).   Hence 
$ \pi_{\phi}$   is a scalar under time coordinate transformations not involving reversal (good enough for Lie differentiation), yielding
\begin{eqnarray} \pounds_{\xi}  \left( \frac{ \sqrt{h} }{N} \dot{\phi } \right)  =  \xi^0 \frac{ \partial} {\partial t}  \left(  \frac{ \sqrt{h} }{N} \dot{\phi } \right). \end{eqnarray}   
Using Hamilton's equations that last expression equals 
$   \xi^0 \frac{ \partial} {\partial t}    \pi_{\phi}. $  

Now we can write the Hamiltonian version of the Killing-like condition for no change:
$$(\exists \xi^0) (\xi^0 \dot{h}_{ij} = 0  \hspace{.1in} \& \hspace{.1in} \xi^0 \dot{\pi}^{ij} = 0 \hspace{.1in} \&  \hspace{.1in} \xi^0 \dot{N} + \dot{\xi^0 } N = 0 \hspace{.1in}  \& \hspace{.1in} \xi^0 \dot{\phi}=0  \hspace{.1in} \& \hspace{.1in}  \xi^0 \dot{\pi}_{\phi} =0).$$ (Naturally $p$ needs to stay $0$ also.) 
Negating to find the condition for \emph{change}, one has 
 \begin{eqnarray}  \lnot (\exists \xi^0) (\xi^0 \dot{h}_{ij} = 0  \hspace{.1in} \& \hspace{.1in} \xi^0 \dot{\pi}^{ij} = 0 \hspace{.1in} \&  \hspace{.1in} \xi^0 \dot{N} + \dot{\xi^0 } N = 0 \hspace{.1in}  \& \hspace{.1in} \xi^0 \dot{\phi}=0  \hspace{.1in} \& \hspace{.1in}  \xi^0 \dot{\pi}_{\phi} =0) \hspace{.1in} \leftrightarrow \nonumber \\
(\forall \xi^0) \hspace{.1in}  \lnot (\xi^0 \dot{h}_{ij} = 0  \hspace{.1in} \& \hspace{.1in} \xi^0 \dot{\pi}^{ij} = 0 \hspace{.1in} \&  \hspace{.1in} \xi^0 \dot{N} + \dot{\xi^0 } N = 0 \hspace{.1in}  \& \hspace{.1in} \xi^0 \dot{\phi}=0  \hspace{.1in} \& \hspace{.1in}  \xi^0 \dot{\pi}_{\phi} =0)  \hspace{.1in} \leftrightarrow  \nonumber \\
(\forall \xi^0) \hspace{.1in}   (\xi^0 \dot{h}_{ij} \neq 0  \hspace{.1in} \lor \hspace{.1in} \xi^0 \dot{\pi}^{ij} \neq 0 \hspace{.1in} \lor  \hspace{.1in} \xi^0 \dot{N} + \dot{\xi^0 } N \neq 0 \hspace{.1in}  \lor \hspace{.1in} \xi^0 \dot{\phi} \neq 0  \hspace{.1in} \lor \hspace{.1in}  \xi^0 \dot{\pi}_{\phi}  \neq 0).
\end{eqnarray}
   The condition $(\xi^0 N),_0 \neq 0$ is  completely unreliable (strongly coordinate-dependent), whereas the other four disjuncts are all invariant.
Hence the  disjunct $(\xi^0 N),_0 \neq 0$ can be dropped.  Vanishing $ \xi^0$ would not count as Killing, so that factor can be dropped.  Now nothing depends on $\xi^0,$ so the quantification over it can be dropped.  Thus the generalized no-Killing condition for change is 
 \begin{eqnarray} \dot{h}_{ij} \neq 0  \hspace{.1in} \lor \hspace{.1in}  \dot{\pi}^{ij}  \neq 0 \hspace{.1in} \lor  \hspace{.1in}  \dot{\phi} \neq 0 \hspace{.1in}  \lor \hspace{.1in} \dot{\pi}_{\phi} \neq 0.
\end{eqnarray}
This is exactly what was derived above from the Hamiltonian time evolution from $H_p.$  Hence the generalized (no) Killing definition of  change agrees with the Hamiltonian definition, as advertised in the title.


\subsection{Lie Derivative from Gauge Generator with Matter Field}

One also wants to be able to implement coordinate transformations in the Hamiltonian formalism, at least using the Hamiltonian equations of motion and  constraints as needed.  With a matter field $\phi$ present, one needs to think about what the new gauge generator $G$ is.  Fortunately one needn't think for long (\emph{c.f.} \cite{CastellaniGaugeGenerator}). There are no new constraints for this very simple matter theory---a feature that wouldn't hold for electromagnetism as the source for gravity, for example.  The gauge generator depends on the primary constraint and the secondary constraint.  The primary is just as before.  The secondary constraint gets a new term, but, crucially, the Poisson bracket `algebra' is unchanged \cite{Sundermeyer}.  Hence one only needs to introduce the modified secondary constraint  $\mathcal{H}_0 + \mathcal{H}_{0\phi}$ in place of $\mathcal{H}_0.$  The result is  
\begin{equation}
G = \epsilon^{\perp}  (\mathcal{H}_0 + \mathcal{H}_{0\phi}) + p\dot{ \epsilon}^{\perp}. 
\end{equation}

One finds that the matter-inclusive gauge generator $G$ has the following Poisson brackets with the canonical variables (re-expressed using $\xi^{0}$ once outside the Poisson bracket): 
  \begin{eqnarray}
\{ h_{ij}, G \} =   \xi^0 \{ h_{ij}, H_p \},           \nonumber \\
\{ \pi^{ij}, G \} =  \xi^0 \{ \pi^{ij}, H_p \},  \nonumber \\
\{ N, G \} = \xi^0 \dot{N} + \dot{\xi^0 } N,    \nonumber \\ 
\{p, G \} =  0  \label{GaugeMatter1}
\end{eqnarray}
as before, 
and now also
 \begin{eqnarray}
\{ \phi, G \} =   \xi^0 \{ \phi, H_p \},           \nonumber \\
\{ \pi_{\phi}, G \} =  \xi^0 \{ \pi_{\phi}, H_p \}.  \label{GaugeMatter2}
\end{eqnarray}

 Using  Hamilton's equations, one has  
\begin{eqnarray}
\{ h_{ij}, G \} =  \xi^0 \dot{h}_{ij},  \nonumber \\ 
\{ \pi^{ij}, G \} = \xi^0 \dot{\pi}^{ij},  \nonumber \\
\{ N, G \} = \xi^0 \dot{N} + \dot{\xi^0 } N, \nonumber \\ 
\{p, G \} = 0, \nonumber \\
\{ \phi, G \} =   \xi^0 \dot{\phi},           \nonumber \\
\{ \pi_{\phi}, G \} =  \xi^0 \dot{\pi}_{\phi}.   
\end{eqnarray}
Excluding $p$, which at least stays $0$, all equations match the Lie derivative formulas appropriate given the transformation properties of the quantities under small change of time coordinate (respectively,  scalar, scalar, weight $1$ density, weight $-1$ density for $p$, scalar, and scalar).  Hence with a massive scalar field present, the gauge generator $G$ does what is expected in this homogeneously truncated stub of General Relativity.
The Hamiltonian definition of change, in addition to being equivalent to the differential geometric definition of change, can be expressed in the differential geometric way using distinctively Hamiltonian resources, which is quite satisfying. 


\subsection{Problem of Space and Many-Fingered Time}

To complete the analysis for GR, one would need, of course, to restore spatial dependence.
While the lack of temporal change in Hamiltonian observables is particularly (in)famous, there is an analogous problem of space, namely, observables are spatially constant as well \cite{TorreObservable}.  Given the relativity of simultaneity, indeed the many-fingered nature of time in GR, temporal change and spatial variation are not naturally separated, except insofar as the Hamiltonian formalism forces the distinction to be made. 

 Fortunately, the answer to the problem of space (missing spatial variation) should be  analogous to the problem of time.  Instead of using the gauge generator for transformations normal to the simultaneity slices (\ref{normalgaugegenerator}), one would use the gauge generator for transformations within the slices (\ref{spatialgaugegenerator}).  The problem of spatially constant observables, like the problem of temporally constant observables, comes from two errors.  First, often there is a failure to maintain equivalence to 4-dimensional differential geometry by failing to treat the $00$ and $0i$ components of the metric (or the lapse and shift) and their conjugate momenta correctly (or by omitting them altogether). Correcting that mistake will yield essential variation with spatial coordinates just in case there are not a suitable number of independent space-like Killing vector fields. Because the Hamiltonian formalism is explicitly spatially covariant, the Hamiltonian version of the story is not very different from the usual differential geometric version, apart from keeping track of the $3+1$ split of 4-dimensional equations for Lie derivatives.  

To get spatial variation in \emph{observables}, one needs to attend to the difference between external and internal gauge symmetries and recognize that demanding $0$ Poisson bracket of observables with the spatial gauge generator (interpreted passively) is the spatial analog of requiring that observables the same at 1 am Eastern  Daylight  Time and 1 am Eastern Standard Time (an hour later).  The gauge generator generates the Lie derivative, which is a transport term that compares two differential space-time points with the same coordinate values in different coordinate systems; that transport term makes all the difference for physical interpretation.  Hence a physically reasonable definition of observables permits spatial variation in just the way that temporal change is permitted.

To get spatio-temporal variation, one uses the full gauge generator, that is, the normal and spatial generators together, with arbitrary descriptor functions (basically a space-time vector field) depending on both space and time. Or one might even forego the $3+1$ split altogether if one has a taste for difficult calculations \cite{AndersonBergmann,Kiriushcheva}.  I do not anticipate any difficulty on this point, but I will not explicitly carry out the outlined steps here.


\subsection{Including More General Matter}

  One would also want to include more general matter fields, including electromagnetism (which introduces  first-class constraints unrelated to GR's), Yang-Mills fields (which add to electromagnetism the complexity of a gauge-covariant rather than gauge-invariant field strength and so more obviously require reckoning with the gauge freedom of the potentials), massive Proca electromagnetism (which has second-class rather than first-class constraints, with nontrivial implications for the gauge generator), and spinor fields, concerning which the ideas of gauge symmetry and Lie differentiation have a large and conflicted literature \cite{PittsSpinor}. In cases where an additional gauge freedom enters, the arguments akin to those above might involve an additional quantification over gauges. To pursue all such matters now would start to submerge the conceptual points about change in GR amidst a longer and more technical discussion.    I expect to address such matters on another occasion, but no particular difficulty seems likely.

%
%

\section{Phase Space-Time and Clarity in Eight Easy Steps}

An issue that arises novelly for General Relativity (with no electromagnetic analog) is that while Hamiltonian techniques are typically applied to phase space, for General Relativity and other theories with velocity-dependent gauge transformations one should use phase space extended by time \cite{MukundaSamuelConstrainedGeometric,SuganoGeneratorQM,SuganoGaugeGenerator,LusannaVelocityHamiltonian}---one might call it phase space-time.   In General Relativity, gauge transformations take the form  $$\pounds_{\xi} g_{\mu\nu} \sim \xi^0 \dot{g}_{\mu\nu} +\ldots,$$ which expression is velocity-dependent.  Hence it is no surprise that something goes wrong in treating GR on phase space.  
Some authors even use a histories formalism for General Relativity, thereby giving  space-time rather than space a still more prominent role \cite{SavvidouHistoriesI,KouletsisHamiltonianHistoryDiss}---though one might hope for a milder revision of the standard phase space formalism with basically the usual Poisson brackets.

Clarity about time evolution in Hamiltonian GR is achieved in eight easy steps that yield Hamiltonian-Lagrangian equivalence:
\begin{enumerate}
  \item Remove the unclarity induced by active diffeomorphisms (which one would have discarded eventually anyway \emph{via} equivalence classes)  in favor of passive coordinate transformations.
  \item Define geometric objects in terms of both coordinates $x^{\mu}$, $x^{\nu\prime},$ and physically individuated points $p$ instead of just $x^{\mu},$  $x^{\nu\prime}.$
  \item Extend the phase space by $x^0$ in view of the velocity-dependent character of the gauge transformations, obtaining phase space-time.
  \item Restore the lapse and shift vector to recover the space-time metric, not just the spatial metric.
  \item Require proper coordinate transformation behavior for the lapse and shift vector somehow or other.  
  \item Recognize that a first-class constraint does not generate a gauge transformation in GR because it violates physical equivalence and  physical law (\emph{c.f.} 4-dimensional differential geometry).
  \item Restore the canonical momenta $p$, $p_i$ conjugate to the lapse and shift vector to the phase space(-time).
  \item Implement 4-dimensional coordinate transformations (at least on-shell) by Poisson bracket with $G$, obtaining 4-dimensional Lie differentiation for solutions of field equations. 
  \end{enumerate}  
Then no confusion remains (except maybe in observables, which I postpone for another occasion). Rather than striving to acquiesce in the mysteries supposedly disclosed by Hamiltonian GR, one can clear them up. Given that such mysteries do not appear in Lagrangian GR, they must be defects in the typical Hamiltonian formalism rather than features of GR.  
 Time evolution in  GR is intricate and in some respects novel, but not bizarre or mysterious---much as with gravitational energy localization  \cite{EnergyGravity}.

One can now see why discussions of reduced phase space for GR have been problematic.  The velocity-dependent character of foliation-changing coordinate transformations implies that phase space-time, not phase space, is the proper arena in which to work.  There just isn't room in phase space to wander off into the future or the past.  
It is thus unclear what it would be to construct a reduced phase \emph{space} for GR \cite{ThebaultCanonicalGRTime} (\emph{c.f.} the description of the process in (\cite{BelotEarman})). Both the original phase space-time and the reduction process need, so to speak, to reach out into the future (and past) in a way that is disanalogous to simpler theories.  
Additional difficulty in constructing a (fully) reduced phase space(-time) for GR arises from the fact, encountered in detail above, that the Hamiltonian formalism implements the equivalent of time-involving coordinate transformations only for solutions of the equations of motion.  
Hence the usual expectation of constructing a space where antecedently recognizably physically equivalent gauge-related configurations that might not be solutions of the equations of motion have been identified as one point, and then formulating dynamics on it, apparently cannot be realized in the usual way.


\section{Relation to Supposedly  Timeless Interpretations of GR}

In view of the claims about the elimination of time that one sometimes encounters in the physics literature, one might wonder about the relation of such claims to the results  above finding real change.  I will take two examples, Rovelli and Barbour.\footnote{I thank an anonymous referee for  a question prompting this section.}  It will appear that their similar rhetoric corresponds to  rather different claims.  My views tend to agree with Rovelli's on many points, whereas I have  reservations about Barbour's reliance on the Baierlein-Sharp-Wheeler Lagrangian density for GR.


\subsection{Rovelli on Time, Space-time, and (Partial) Observables}

Rovelli's work tends to appear with a rhetorical flourish suggestive of distinctive and dramatic claims.  One is urged to ``Forget Time'' \cite{RovelliForgetTime} and to complete the ``unfinished revolution''  of GR in quantization \cite[p. 3]{RovelliBook}, for example.    Many of Rovelli's rhetorically dramatic points and innovations amount, on further inspection, to stone-cold-sober observations about the need to maintain equivalence to Lagrangian GR in the face of entrenched habits of violating it in the canonical-focussed quantum gravity community.  Hence my conclusions actually substantially overlap in both substance and motivation with Rovelli's.  I will paint with a broad brush, leaving room for further elaboration elsewhere on some points.  

Forgetting time, as practiced by Rovelli, seems to mean, at least in part, ``remember space-time.''  If one attempts to formulate GR on phase space (not phase space-time) as if often done, then one has in effect forgotten time by singling it out and then dropping it altogether. Why think that a theory of space-time could be formulated without time as part of the mathematical space?  By contrast, Rovelli's vision of forgetting time prohibits singling time out.  Such a vision forbids both including time on special footing (which seems to be the main advertised target) and singling time out  and then dropping it (which often happens).   What he  motivates with picturesque rhetoric is largely what has been urged above  \cite{MukundaSamuelConstrainedGeometric,SuganoGeneratorQM,SuganoGaugeGenerator,LusannaVelocityHamiltonian}  on such nuts-and-bolts  mathematical grounds as  velocity-dependent gauge transformations in GR.  

Whether one engages in the GR-advocacy of completing its ``unfinished revolution'' in quantization or not, 
it does seem evident that quantization  of GR will be effected  better, if more of the distinctive classical features of GR are preserved under quantization.  Demanding of any canonical quantization program that it be based upon a classical Hamiltonian formalism equivalent to the Lagrangian (apart from inevitable topological issues) is one important means to that end, at least if canonical quantization is worth trying (as it surely is).  Hence completing the unfinished revolution of GR seems to mean, at least in part, that one should preserve Hamiltonian-Lagrangian equivalence while using a Hamiltonian and contemplating quantization.  

Regarding observables, Rovelli has found the traditional notion of observables sufficiently unwieldy to use as to develop a weaker notion of ``partial observables.''  He emphasizes physical/relational point individuation, as opposed to the primitive individuation introduced in modern-style differential geometry \cite{RovelliObservable,RovelliPartialObservables}.  ``General relativistic systems are formulated in terms of variables\ldots that evolve with respect to each other. General relativity expresses relations between these, but in general we cannot solve for one in terms of the others.  Partial observables are genuinely on the same footing.'' \cite{RovelliPartialObservables}. Coordinates can be eliminated from observations \cite{RovelliCovariantHamiltonian}.   One needs 5 scalar fields in order to observe 1, because 4 are used up in identifying the relevant space-time points by inversion. Traditionally, such considerations were already implemented in classical differential geometry by its avoidance of primitive point individuation (suited to Einstein's point-coincidence argument) and its relativization of physical quantities to a set of arbitrary conventions for labeling space-time (coordinates).  Alternatively, one could use {\bf in}variant but non-numerical tensors-in-themselves ${\bf g} = g_{\mu\nu} {\bf d}x^{\mu}\otimes{\bf d}x^{\nu}$ as operators waiting to give a number relative to a basis; then one needs to reckon with the fact that physics has traditionally been set up for numerical entities and take appropriate measures.  Rovelli's partial observables bear a strong  resemblance to what observables always should have been in Hamiltonian GR, and what they implicitly were in Lagrangian GR.

In short, much of Rovelli's project involves understanding GR correctly in the face of inadequate efforts to do so.  Many of his key themes are quite helpful for uncovering real change---at least $B$-series change, different properties at different times---in GR.


\subsection{Barbour's Timelessness of GR and the Fragile BSW Action}

Barbour has long argued for the elimination of time from GR \cite{BarbourTimeless1,BarbourTimeless2}.  A cornerstone of his interpretation is the possibility of eliminating the lapse $N$ from the action on GR, as performed long ago by Baierlein, Sharp and Wheeler (BSW) \cite{BSW}.  Noting the fragility of the BSW procedure  gives one pause about relying too heavily on it for conceptual purposes.

Barbour's claims are both bold (in content and form) and BSW-dependent. In the second paper he writes about the first: ``I have demonstrated that there is a precise sense in which classical general relativity (GR) is timeless and frameless.''  He continues a few pages later:
\begin{quote}
Indeed, the dilemma in quantum gravity is to decide whether classical GR's remarkable four-dimensionality (hitherto regarded as its defining characteristic) is more fundamental than the structure of geometrodynamics as a timeless theory in a relative configuration space ([1], section 14).  General relativity is like a Shakespearian drama with subplot so prominent one might confuse it with the main plot.  Is Gloucester more than Lear?  Is foliation invariance, the substance of spacetime, more than ephemeris time and intrinsic equilocality?

My answer is the BSW Lagrangian (35) of [1].  I believe this is a deeper expression of what GR is about than the original Hilbert-Einstein 4-action, since it is so economic with basic variables:  no lapse, the shift is an auxiliary equilocality shuffler\ldots. 

The BSW Lagrangian exhibits a structural hierarchy:  without the square root and the equilocality shuffler it all dissolves.  \emph{They} are the dynamical framework of GR.\ldots Foliation invariance is the \emph{subplot.}   \cite[p. 2878]{BarbourTimeless2}. \end{quote}
With so much interpretive weight resting on the BSW Lagrangian for GR, one should attend to how generally it exists.  
The BSW elimination of $N$ happens at the Lagrangian level, where the velocities  
$\dot{h}_{ij}$, not any canonical momenta, are present. 
Whether a quantity (such as $N$) is an auxiliary field---whether it enters the action in an essentially nonlinear and algebraic way, in the simplest case, though one might tolerate spatial derivatives---depends on what the other quantities in the problem are.  Hence $N$ appears nonlinearly in the Lagrangian formulation (where the 3-metric $h_{ij}$  and its velocity appear), but linearly in the Hamiltonian (where the 3-metric and its canonical momentum appear):   the transition from the velocity to the momentum has swallowed up some of the $N$-dependence due to the $N$-dependent relation between $\pi^{ij}$ and $\dot{ h }_{ij}$ \cite{MTW,Wald}.   Hence a simple statement that the lapse is (or isn't) an auxiliary field is ill-defined without more context.  
 But the dependence on $N$ in the Lagrangian density is  indeed essentially nonlinear and algebraic:
\begin{equation}
\mathcal{L} = \frac{  \sqrt{h} }{4N} (h^{ac}h^{bd} -h^{ab} h^{cd}) \dot{h}_{ab} \dot{h}_{cd} + N \sqrt{h} R + \ldots,
\end{equation} where the omitted terms involve the shift vector and $R$ is the \emph{spatial} Ricci scalar \cite{MTW,Wald}. BSW eliminate $N$ by solving $\frac{ \partial \mathcal{L} }{\partial N}=0$ for $N$. Continuing to omit the terms with the shift for simplicity, one has $\sqrt{h}R = N^{-2} \frac{ \sqrt{h} }{4} (h^{ac}h^{bd} -h^{ab} h^{cd}) \dot{h}_{ab} \dot{h}_{cd},$ which can be solved for $N^2$ \emph{only if } $R\neq 0.$   But what if $R=0$?  Then one cannot solve for $N$; BSW's equation $7$ cannot be obtained.  Purely inverse quadratic dependence on $N$ in $\frac{\partial \mathcal{L} }{\partial N}$ (with  no $N^0$ term) makes $N$ in effect a Lagrange multiplier (linear dependence on $N^{-2}$), not an auxiliary field (essentially nonlinear and hence soluble dependence on $N$). Something bad happens at moments of time symmetry as well, such as if an expanding universe stopped and started contracting: $N$ disappears from  $\frac{ \partial \mathcal{L} }{\partial N}=0.$ Whereas my treatment of Bianchi I cosmological models preserves most of the relevant features of GR (apart from many-fingered time) and has no difficulty finding time and change, the BSW lapse-elimination on which Barbour's time-eliminating interpretation is not even available.   It seems perhaps unwise to place too much reliance on a manipulation that does not always work.  At any rate the fragility of the BSW procedure makes Barbour's interpretation  involving the elimination of time less than compelling.

  Furthermore, one might  wish to formulate GR in a way that doesn't exclude in advance any possibility of perturbative treatment.  One can then ascertain which features of the theory exist even without interactions (the limit as Newton's $G$ vanishes), and which depend on interactions.  To that end one can make a perturbation-friendly change of variables, which includes splitting off the ``vacuum expectation values'':  $h_{ij}= \delta_{ij} + \sqrt{32\pi G} \gamma_{ij}$, $N=1 + \sqrt{32 \pi G}  n$ and rewriting using $\gamma_{ij}$ (at least in derivatives) and $n$.  In the process one also gives up the now-usual unphysical normalization of the ADM Lagrangian \cite[p. 520]{MTW} in favor of the physically standard normalization $\mathcal{L} = \frac{1}{2} \dot{\gamma}_{ij}^2 \ldots$.    Applying the well-known possibility of splitting the weight $1$ Ricci scalar density to three dimensions ($\sqrt{h}R= \,^{3}\Gamma\Gamma +div$), suppressing all terms involving the shift, and dropping all terms independent of $n$, one has these essential BSW-related terms in $\mathcal{L}$ pertaining to the possibility of eliminating $N$ by its equation of motion:
\begin{equation} \underbrace{ \frac{ n\cdot div}{\sqrt{8 \pi G}} }_{O(\sqrt{G}^0)} +  \underbrace{n \left( \frac{  \,^3\Gamma\Gamma }{\sqrt{8 \pi G}}   -\sqrt{8 \pi G h} \dot{\gamma}_{ab} \dot{\gamma}_{cd} (h^{ac} h^{bd} - h^{ab} h^{cd})\right)}_{O(\sqrt{G})}   + \underbrace{ n^2 16 \pi G \sqrt{h} \dot{\gamma}_{ab} \dot{\gamma}_{cd} (h^{ac} h^{bd} - h^{ab} h^{cd}) }_{O(\sqrt{G}^2)} -\ldots. \end{equation}  

One now sees an additional sense in which the BSW elimination of $N$ is fragile:  it works not at zeroth order in the coupling constant $\sqrt{32 \pi G}$ (the free or linearized theory), nor even at first interacting order in  $\sqrt{32 \pi G}$,  but finally at second interacting order $O(\sqrt{G}^2)$ before one can even solve at all for $N$---to say nothing of solving and getting anything at least approximating the perturbation-unfriendly BSW answer!  At zeroth order in $\sqrt{G}$, $n$ is  a Lagrange multiplier implying a Poisson-like constraint equation.  At first order in $\sqrt{G}$ (which is still linear in $n$), $n$ is a Lagrange multiplier for a constraint equation expanded to include energy terms. Finally at second order in $\sqrt{G},$ which is the lowest order in which $n^2$ appears, one can vary $n$ for an equation that can be solved for $n$, which is essential to the BSW trick. 
If one solves that equation for $n$, one still doesn't get anything like the `right' BSW answer for $n$ and hence $N$, however.  The same should hold at higher orders $n^3$, $n^4,$ $n^5$ (soon leaving the realm where the equation can be solved in closed form as one gets a quintic polynominal), \emph{etc.}   One has to work to all orders and then reverse the geometric series expansion. Hence the BSW lapse elimination depends as strongly as it possibly could on interaction terms.  Related points were made  prior to BSW by Peres and Rosen \cite{PeresRosenCauchyI}. 
 It is  unusual for the essence of a theory to be completely absent in the two lowest order terms (which already account for much of the theory's empirical success), and then have a sequence of partly mistaken essences for every higher perturbative order (which will quickly account for any remaining empirical confirmation \cite{BlanchetNonmetric}), with the true essence becoming appearing only nonperturbatively (and far transcending empirical test).   One might see this fact as reason to wonder whether the BSW action really uncovers the essence of GR.  Because all of our evidence for GR requires less than exact nonperturbative treatment, that evidence would fit rival theories that approximate GR to some high order in $\sqrt{G}$ but not all orders, and which therefore lack the true BSW essence of GR.

 Whereas Barbour has wanted to see how empty the glass of time in GR is, I find it reasonable  to view the glass as mostly full. To that end his doubts about aspects of the usual meanings attributed to first-class constraints are  useful \cite{BarbourFosterPrimary}.


\section{Resolution of Earman-Maudlin Standoff }

Successfully finding change in Hamiltonian GR  resolves the standoff between Earman and Maudlin \cite{EarmanMcTaggart,MaudlinMcTaggart}.  
Earman invoked Hamiltonian GR and inferred no real  change (more or less---one is left contending with a neologistic ``D series'' in the spirit of McTaggart and delving into the philosophy of mind), whereas Maudlin invoked common sense's  real change and rejected Hamiltonian GR.
Earman's enthusiasm for Hamiltonian GR is not misplaced, but various errors commonly found in the physics literature needed to be corrected.
 Earman's mere ``surface structure'' of tensor calculus and the presence or absence of Killing vector fields in fact  provided the key:  it gives same answer as the corrected Hamiltonian treatment.
His ``deep structure'' is erroneous in ways that I  discuss here, the prequel \cite{FirstClassNotGaugeEM} and its GR-oriented companion (in preparation \cite{FirstClassNotGaugeGR}), and the sequel on observables  \cite{Observables}.
 Maudlin's common-sense detection of change is vindicated, but his dismissal of the Hamiltonian formalism isn't.  Instead a Lagrangian-guided reform of the Hamiltonian formalism has been achieved, the continuing need for which (extending conceptually what Mukunda, Castellani, Sugano, Kimura, Pons, Salisbury, Shepley, Sundermeyer and a few others have done technically) might not yet have been recognized without Maudlin's critique.  

Of course requiring that the Hamiltonian formalism  agree with the Lagrangian formalism simply had to work.
 Change in Hamiltonian GR is neither distinctively  canonical nor problematic, at any rate not classically. Presumably there are quantum consequences, but I do not attempt here to say what they are. It might be that resolving problems in classical canonical GR simultaneously loosens the presumed connections between classical and quantum canonical GR that Bergmann expected. 
 Enforcing the equivalence of the unclear constrained Hamiltonian formalism with the clear Lagrangian formalism is  evident in early works by Bergmann and collaborators (\emph{e.g.},  \cite{AndersonBergmann}).  Eventually Bergmann took shortcuts about observables, as Dirac did about gauge transformations and even the size of the phase space.  Recovering Hamiltonian-Lagrangian equivalence removes confusion.
 Real change has been found without delving much into the ``observables'' thicket, though addressing such issues in a Lagrangian-equivalent way can bring additional clarity \cite{Observables}.

\section{Philosophical Accounts of Time and Change in GR}

There isn't any conceptual difficulty in locating change in General Relativity, either in the Lagrangian formalism or in the Hamiltonian formalism, once the latter is set up properly.  The former claim is fairly widely accepted.  The latter differs from widely shared views among philosophers \cite{BelotPoP,BelotEarman,BelotEarmanButterfield,HuggettWuthrichTimeQG,RicklesTimeStructureQG}, where the idea  that a first-class constraint generates a gauge transformation has been widely influential.

 Demanding and successfully achieving a Hamiltonian formalism equivalent to the Lagrangian one will ensure that change is equally discernible both places.  But much of the conceptual reflection on GR and much of the effort to quantize it canonically have been carried out with Hamiltonian formalism(s) inequivalent to the Lagrangian.  A main root  of the difficulty is the doctrine that a first-class constraint generates a gauge transformation.   That doctrine, widely assumed by high authorities in theoretical/mathematical physics \cite{HenneauxTeitelboim,GotayNesterHinds} and already apparent in the work of Dirac and Bergmann in the 1950s (but not in (\cite{AndersonBergmann}), which employs $G$), is nonetheless readily falsifiable by direct calculation using the example of electromagnetism \cite{FirstClassNotGaugeEM}.
This doctrine is often associated with the extended Hamiltonian, Dirac's idea that one may add all the first-class constraints (not just first-class primary constraints) to the Hamiltonian with arbitrary coefficients, and needs to do so to exhibit the full gauge freedom \cite{DiracLQM}.  
Instead, a gauge transformation is actually generated by a \emph{special combination} of first-class constraints, the gauge generator $G$ of Anderson and Bergmann, which disappeared for about 30 years and started resurfacing slowly in the 1980s  \cite{MukundaGaugeGenerator,CastellaniGaugeGenerator,SuganoGaugeGenerator,GraciaPons,SuganoExtended,PonsSalisburyShepley,ShepleyPonsSalisburyTurkish,PonsSalisbury}.  The gauge generator $G$ is associated with the primary Hamiltonian, which adds to the canonical Hamiltonian only the primary first-class constraints.  The primary Hamiltonian is equivalent to the Lagrangian, whereas the extended Hamiltonian is not.  Proponents of the extended Hamiltonian claim that it is equivalent to the Lagrangian for observable quantities, the difference being additional extra gauge freedom \cite{HenneauxTeitelboim}.  That turns out not to be the case once one is clear about primordial observables \emph{vs.} auxiliary fields like canonical momenta \cite{FirstClassNotGaugeEM}.   The canonical momentum conjugate to $A_i,$ far from being the primordial observable electric field as is sometimes claimed, is just an auxiliary field that one can integrate out using its algebraic equation of motion due to its nonlinear algebraic appearance in the Hamiltonian action $S=\int dt d^3x (p \dot{q}-\mathcal{H}).$ By such moves one thereby recovers the Lagrangian action $\int dt d^3x \mathcal{L},$ which gets by just fine without the conjugate momenta.  $A_i$ and $A_0$ are what couple to current and charge densities, so they (through a gauge-invariant combination of their derivatives \cite{Jackson}) constitute the primordial observable electric field. 


\subsection{Comparison to Belot and Earman}

Belot and Earman's treatments, being especially thorough, early and influential, will repay  detailed attention. 
One should note that, at least on technical grounds, Earman is perhaps of two minds,  assuming that a first class constraint generates a gauge transformation, but also briefly exhibiting the gauge generator \cite{EarmanOde}. But the former view is clearly dominant in his work.

It appears that the difference between my conclusions about the ease of finding change in GR and Belot's conclusions about the  difficulty of finding change in GR are caused largely by the divide between my adopting the Lagrangian-equivalent primary Hamiltonian and the specially tuned sum of first-class constraints $G$ as generating gauge transformations on the one hand, and Belot's and Earman's adopting  the Lagrangian-inequivalent extended Hamiltonian and an arbitrary sum of first-class constraints as generating gauge transformations on the other \cite{BelotEarman}.\footnote{ I thank Oliver Pooley for very helpful discussion on this matter.}


While much of the difficulty in finding change arises in Hamiltonian treatments, Belot finds difficulty already at the Lagrangian level \cite[p. 172]{BelotPoP}, presumed to be due to the inclusion of unphysical variables.  ``The remedy is reduction --- the reduced space of solutions and the reduced space of initial data are symplectic and isomorphic.''  While there is some sense in which reduction in the loose sense of looking for a small nice thing related to a large messy thing perhaps needs to be the answer, constructing a new smaller and more abstract space is not the only option.  An alternative, as was displayed above, is to work in terms of a specific solution (or, for that matter, a solution and all its gauge-transforms) and apply a gauge-invariant test for the \emph{existence} of a special gauge (in vacuum GR, a coordinate system) in which time dependence is absent. One does not need to find that coordinate system in advance of recognizing its specialness; indeed one does not need to find it at all (if it exists).   One only needs to ascertain whether the Killing equation $\pounds_{\xi} g_{\mu\nu}=0$ has a solution for a time-like vector field $\xi^{\mu}$, a gauge-invariant statement (true in all coordinate systems),  to learn whether the metric lacks change.  Likewise for matter, at least assuming that the matter fields are geometric objects and do not have additional gauge groups (hence there are further technical complications for GR + electromagnetism or GR + spinors).  If neither the metric nor any matter changes, then there is no change; if something changes, then there is change.  This is a straightforward question of classical tensor calculus, requiring none of the sophistication involved in reduction. In effect one ascertains whether one \emph{could} gauge-fix with a certain result, without actually gauge-fixing. The test for change is local---there can be a time-like Killing vector in some places but not others (witness the Schwarzschild solution) and hence does not require the distinction between cosmological and asymptotically flat solutions.  The use of a gauge-invariant test for a `good' gauge requires the use of quantifiers, as appeared above. 

   Many of Belot's  remaining difficulties relevant to the Lagrangian formalism are consequences of seeking Hamiltonian-Lagrangian equivalence from the wrong end---that is, accepting widely asserted  features of the Hamiltonian formalism  as true and expecting the Lagrangian to match, rather than asking that  the Hamiltonian  match the more perspicuous Lagrangian. Some of these supposed features have to do with constrained Hamiltonian theories with first-class constraints in general, such as too broad a view of what generates a gauge transformation.  Others have to do with peculiarities of GR, for which the gauge symmetry is external (motivating covariance rather than invariance), the gauge transformations normal to the simultaneity hypersurface only relate physically equivalent and physically meaningful states \emph{on-shell} (using equations of motion) because  $\mathcal{H}_0$ is quadratic in momenta \cite{FradkinVilkoviskyHLEquivalence,CastellaniGaugeGenerator,MukhanovWipf}, the velocity dependence of the gauge transformations necessitates extending phase space by time,  \emph{etc.}.  The very phrase ``reduced phase space'' has erroneous presuppositions.   It is also by no means clear (\emph{c.f.} p. 211) that Lagrangian quantization ought to proceed in terms of true degrees of freedom---witness the vast and sophisticated literature for quantization of GR that leaves the gauge freedom in.


Before delving into the challenges of canonical GR, it is advisable to look at the standard test bed theory, Maxwell's electromagnetism, as presented in Belot's version of a constrained Hamiltonian formalism \cite{BelotElectromagnetism,BelotGauge}.  One first notices the omission not merely of the canonical momentum $p^0$ conjugate to the scalar potential $A_0$ (which is common enough and not automatically disastrous), but also the omission of $A_0$ itself, from the formulation.  Is this elimination achieved by gauge fixing $A_0=0$ \emph{via} the Dirac bracket, the usual gauge-fixing technology of Dirac-Bergmann constrained dynamics \cite{Sundermeyer}?  Such a formalism would be possible, though not trivial; \emph{e.g.}, what is the canonical generator of whatever gauge transformations might remain \cite{MukundaGaugeGenerator}?  It would be awkward to couple Belot's Hamiltonian electromagnetism to charge without $A_0,$ because $A_0 \rho$ is the standard interaction term with charge density.  But difficulties exist even for the vacuum case.  One dilemma pertains to the origin of the Gauss-like constraint $\nabla \cdot \vec{E} = 0.$
(I have here employed Belot's notation of using $E$ for the canonical momentum.  Such a notation in fact is a temptation to conflate entities that are quite distinct in their gauge transformation properties, $F_{0i}$ (a familiar function of derivatives of $A_{\mu}$) and the canonical momentum.  These quantities are in fact not even related until one uses the equations of motion:  $p$ is independent of $\dot{q}.$  That is why I called the constraint Gauss-\emph{like}.)    
The Hamiltonian is given as (simplifying to flat space and Cartesian coordinates for simplicity) $$\int d^3x \frac{1}{2} (\vec{E} \cdot \vec{E} +      |\nabla \times \vec{A}|^2).$$

 Now one faces a dilemma.
If the condition  $\nabla \cdot \vec{E} = 0$ is not employed in the variational principle, then how does it arise later?   Note that gauge-fixing $A_0=0$ in the usual treatment would require that any lingering gauge transformations be time-independent in view of the usual gauge freedom $\delta A_{\mu} = -\partial_{\mu} \epsilon.$  Thus Belot's notation $\Lambda(t)$ is misleading in implying time dependence of the descriptor $\Lambda$ (basically my $-\epsilon$).  Note also that, in contrast to the more usual formulation of electromagnetism that includes $A_0$, Belot's alleged gauge transformation descriptor $\Lambda$ doesn't cancel out of the Maxwell equation $\dot{A}=-\vec{E}$ under the alleged gauge transformation $\vec{A}\rightarrow \vec{A} + \nabla \Lambda$---not unless one discards the advertised time dependence (or the spatial dependence) of $\Lambda(t)$ (called $g(t)$ in the later paper). One also spoils the Gauss-like law  $\nabla \cdot \vec{E}=0$ unless $\nabla^2 \dot{g}=0.$ Without time-dependent arbitrary functions, by definition there is no gauge freedom.   Thus there actually is no gauge freedom in Belot's formulation, and no threat of indeterminism.

 On the other hand, suppose that $\nabla \cdot \vec{E} = 0$ is imposed secretly in the variational principle.  That would explain why the constraint exists later.  But there is a cost in the variational principle, which must  make use of a transverse-longitudinal decomposition (\emph{e.g.}, \cite{DeserMasslessLimit,Marzban}).  Then one gets not $\dot{\vec{A}}=-\vec{E},$ but 
only its transverse part $$\frac{\partial}{\partial t}\left(\vec{A} - \frac{\nabla (\nabla \cdot \vec{A})}{\nabla^2}\right)= -\vec{E}:$$  the gradient part is projected away, leaving only what is spatially divergenceless.  This expression is perhaps most readily motivated using spatial Fourier transformations, and can also be understood in terms of Green's functions.  This transverse projection ensures that $ \nabla \Lambda$, the supposed gauge transformation, is immediately eliminated by the projection operation.  Resorting to Fourier space or inverting a differential operator using a Green's function is of course a spatially \emph{nonlocal} operation.  Such a blatant threat to locality will pose grave difficulties for the project of understanding senses of locality, the Aharonov-Bohm effect, \emph{etc.} later in the papers.


Turning to Belot's article on time and change in mechanics generally and in GR in particular, one finds that section 3.3 \cite{BelotPoP} on presymplectic manifolds presents a collection of stipulative definitions, some of them about familiar words like ``gauge,'' some of them less familiar like ``presymplectic'':  ``We call,'' ``We define,'' \emph{etc.}  While clearly the definitions of long-familiar words are intended to be roughly equivalent to more traditional ones like $A_{\mu} \rightarrow A_{\mu} - \partial_{\mu} \epsilon(t,x)$ \cite[p. 189]{BelotPoP}, that equivalence is not shown---not there, and not successfully elsewhere (\emph{e.g.} \cite{GotayNesterHinds,HenneauxTeitelboim}) either. 
Beneath the surface, evidently, is the doctrine that a first-class constraint generates a gauge transformation \cite{BelotEarman}---which, alas, isn't true \cite{FirstClassNotGaugeEM}.  Gauge transformations for electromagnetism make sense ``off-shell'' (without using any of Hamilton's equations), and there just isn't any relationship at all between the canonical momentum and the electric field (a function of derivatives of $A_{\mu}$---the field that couples to charge density in the term $A_{\mu} \mathcal{J}^{\mu}$) in that context.   Hence preserving the magnetic field (the curl of the 3-vector $A_m$) and the canonical momentum, Belot's necessary and sufficient conditions for physical equivalence \cite[p. 189]{BelotPoP}, is  necessary but \emph{not} sufficient.  One also needs to preserve the electric field, which equals the canonical momenta (up to a sign) only \emph{on}-shell \emph{via} $\dot{q} = \frac{\delta H}{\delta p}$.  (Belot and Earman also present the electric field as though it were itself the canonical momenta \cite{BelotEarman}.) That equality is spoiled by arbitrary combinations of first-class constraints; it is preserved only by the specially tuned combination $G$.  Indeed one can derive the form of $G$ by requiring that the change in the electric field from the primary constraint and the change in the electric field from the secondary straight cancel out \cite{FirstClassNotGaugeEM}. The relation between the canonical momentum and the electric field comes from calculating a Poisson bracket and so cannot be used inside another Poisson bracket (such as in calculating a gauge transformation).

Turning to General Relativity itself, one finds both analogs of the issues for electromagnetism and new ones.  
One issue that becomes important is whether one uses  active diffeomorphisms (as Belot does) or  passive coordinate transformations.  Active diffeomorphisms presuppose that space-time points are individuated mathematically independently of what happens there. But the lesson of Einstein's point-coincidence argument seems to be that space-time points are individuated physically by virtue of what happens there \cite{HoeferHole}.  
It seems to start off on the wrong foot to presuppose a mathematical formalism ill-adapted to the lessons of Einstein's point-coincidence argument 
by individuating space-time points primitively for mathematical purposes, and then to repudiate the physical meaning of that individuation.  That curious piece of mathematical metaphysics  lacks two merits of coordinate systems, namely, being descriptively rich enough to do tensor calculus, and being manifestly just one of many equally good options and hence less tempting to take with undue seriousness.  Geographers and classical differential geometers have rules for changing coordinate systems, but it seems awkward to change names of individuals, or to introduce individuals and then annihilate or conflate them.  Clarity about passive coordinate transformations will play a role below in identifying gauge transformations and hence gauge-invariant quantities.  

A second issue that arises novelly for General Relativity, as noted above, is the velocity-dependent gauge transformations and consequent need to  use phase space extended by time.  While one can attempt to carry over to GR a reduced phase space construction that works for theories with internal velocity-independent transformations \cite{BelotEarman}, the results are very unlikely to mean what one hoped.    This fact relates to the admission elsewhere in the paper that it isn't obvious how changes of time coordinate are implemented in their formalism \cite{BelotEarman}. 
Indeed when one does implement changes of time coordinate (as discussed above), one doesn't  know which points in phase space-time are physically equivalent under the gauge transformations that (on-shell!) change the time coordinate until \emph{after} the equations of motion are used.  Hence the usual idea of reduced phase space as implementing dynamics on a space where physically equivalent gauge-related descriptions have been \emph{identified in advance} is impossible.  One would therefore need to rethink reduced phase space-time from the ground up for GR, as far as changes of time coordinate are concerned, before philosophizing about it.

A third issue involves what to do with the less interesting field components and their momenta.  Although Belot now keeps the electric scalar potential $A_0$ (in contrast to \cite{BelotElectromagnetism}), for GR he gives short shrift to the lapse and the shift vector, the 40\% of the space-time metric by which it transcends the spatial metric.  He apparently discards the lapse and shift in choosing a Gaussian slicing \cite[p. 201]{BelotPoP}.  Belot and Earman discard the electric scalar potential  and claim to follow Beig \cite{BeigBadHonnef} in eliminating the lapse and shift \cite{BelotEarman}.  For Beig that means only keeping them as freely prescribed functions of space and time without the Hamiltonian apparatus of canonical momenta and Hamilton's equations; it might mean something stronger for Belot and Earman. 
  Losing some of the $q$'s, besides leaving one unable to infer the space-time metric, makes it well-nigh impossible to express 4-dimensional coordinate transformations. 

To discuss further points at issue particularly in GR, it is best to quote the end of p. 201 and much of p. 202  for contrast.
\begin{quote}
 The gauge orbits of [the presymplectic form] $\omega$ have the   following structure: initial data sets $(q, \pi)$ and $(q^{\prime}, \pi^{\prime})$ belong to the
same gauge orbit if and only if they arise as initial data for the same
solution g.[Footnote suppressed]

2. \emph{Construct a Hamiltonian}. Application of the usual rule for constructing
a Hamiltonian given a Lagrangian leads to the Hamiltonian $h \equiv 0.$  

3.  \emph{Construct dynamics}.   Imposing the usual dynamical equation, according
to which the dynamical trajectories are generated by the vector field(s)
$X_h$ solving $\omega(X_h, \cdot) = dh$, leads to the conclusion that dynamical trajectories
are those curves generated by null vector fields. So a curve in
$\mathcal{I}$ [the space of initial data] is a dynamical trajectory if and only if it stays always in the same
gauge orbit. This is, of course, physically useless —-- since normally
we expect dynamical trajectories for a theory with gauge symmetries
to encode physical information by passing from gauge orbit to gauge
orbit. But in the present case, nothing else could have been hoped
for. A non-zero Hamiltonian would have led to dynamical trajectories
which passed from gauge orbit to gauge orbit --— but this would have
been physical nonsense (and worse than useless). For such dynamics
would have carried us from an initial state that could be thought of
as an instantaneous state for solution $g$ to a later instantaneous state
that could not be thought of as an instantaneous state for solution $g$.
In doing so, it would have turned out to encode dynamical information
very different from that encoded in Einstein's field equations. \cite[pp. 201, 202]{BelotPoP}
\end{quote} 

The claim that the Hamiltonian vanishes  \emph{identically}  (also in \cite{BelotEarmanButterfield}) is not correct.  The  method for constructing a Hamiltonian in constrained dynamics gives a Hamiltonian that is only weakly equal to 0 (plus boundary terms, which do not matter) \cite{Sundermeyer}; the gradient is not $0$, so Poisson brackets with the Hamiltonian need not be $0$.  Weak equality is by construction compatible with non-zero Poisson brackets and hence a non-zero Hamiltonian vector field. Belot and Earman point to chapter 4 of (\cite{HenneauxTeitelboim}).  There one finds the following:  
\begin{quote} If the $q$'s and $p$'s transform as scalars under reparametrizations, the $p\dot{q}$-term in the action transforms as a scalar density, and its time integral is therefore invariant by itself.\ldots  
\emph{Thus, if $q$ and $p$ transform as scalars under time reparametrizations, the Hamiltonian is {\bf (weakly)} zero for a generally covariant system.} \cite[pp. 105, 106, italics in the original, but boldface added]{HenneauxTeitelboim} \end{quote}
 As appeared above in the flurry of Lie derivatives, the relevant $q$'s and $p$'s     are scalars under time reparametrization.  (The lapse fits in with Henneaux and Teitelboim's $u$'s, which are densities, as is the lapse.) 
In GR the Hamiltonian is weakly $0$ (apart from possible boundary terms), but it nonetheless has nonzero Poisson brackets and so is not prohibited from generating real time evolution.

The primary Hamiltonian leads to equations equivalent to Einstein's equations \cite{Sundermeyer}.  Such dynamics takes one set of initial data to another set of (what one could regard as initial) data with (typically) different properties---the universe has expanded, for example, or gravitational waves have propagated, or some such. Change has occurred. Both moments are parts of the same space-time.  Gauge transformations can be divided into purely spatial ones and those changing the time as well.  Purely spatial ones, generated by $G[\epsilon^i, \dot{\epsilon}^i]$ depending on a spatially projected 3-vector and its velocity, take each moment under one coordinate description into that same moment under another coordinate description  with the same time coordinate but different spatial coordinates. Coordinate transformations involving time cannot be implemented on phase space, but live rather on phase space extended by time, because they are velocity-dependent.  Such a gauge transformation, which essentially involves the normally projected gauge generator $G[\epsilon, \dot{\epsilon}]$, acts on an entire space-time (trajectory, history) and repaints coordinate labels onto it (or at any rate acts in that way in the overlap of the two relevant coordinate charts); the relabeling happens on a 4-dimensional blob, not a 3-dimensional one.  Relabeling a space-time with coordinates leaves one on the same gauge orbit, but that in no way implies the absence of change.  Change is indicated by nonzero Lie derivatives with respect to all time-like vector fields.

The representation of changeable quantities proves to be rather harder for Belot than it is on my view. He surveys various spaces on which one might try to represent changeable quantities---the space of solutions, the reduced space of solutions, and the reduced space of initial data  \cite[pp. 203, 204]{BelotPoP} and comes up empty every time.   
\begin{quote} And on the space of initial data we face an unattractive dilemma:  if we seek to represent changeable quantities by non-gauge invariant functions, then we face indeterminism; if we employ gauge-invariant functions, then we are faced with essentially the same situation we met in the reduced space of initial data. \cite[p. 204]{BelotPoP}
\end{quote}
 At this point it becomes important to employ exclusively passive coordinate transformations in order to have a clear idea of what gauge-invariance is and hence what the gauge-invariant functions are.  Thus gauge transformations are coordinate transformations.  Famously, scalars are coordinate-invariant.  Hence scalars that depend on time exhibit change; an independent set of Weyl curvature scalars to form a coordinate system is one example.  Quantities that aren't scalars (gauge-invariant) might still be geometric objects (gauge-\emph{covariant})---contravariant vectors, covariant vectors, various kinds of tensors, tensor densities, \emph{etc.}, the meat and drink of classical differential geometry \cite{Schouten,Anderson}.  Lie differentiation is the tool to ascertain time dependence.  In cases where (there exists a coordinate system such that) the metric is independent of time, one has a time-like Killing vector field: there exists a vector field $\xi^{\alpha}$ such that   $\pounds_{\xi} g_{\mu\nu}=0$ and $\xi^{\alpha}$  is time-like.  The components $g_{\mu\nu}$ are not gauge-invariant, but they are gauge-covariant, which is good enough.  (The metric-in-itself ${\bf g}(p)=  g_{\mu\nu} {\bf d}x^{\mu} \otimes {\bf d}x^{\nu}$ is gauge-invariant, but the usual modern definition of Lie derivatives involves active diffeomorphisms; the geometric object $\{ g_{\mu\nu} \}$ (in \emph{all} coordinates \cite{Nijenhuis,Schouten,Anderson,Trautman}) is also   gauge-invariant, but dealing with every coordinate system at once introduces needless complication.  Hence the classical approach of using any arbitrary coordinate system as representative is attractive.)   It makes no difference whether the space-time is spatially closed, asymptotically flat, or neither, because change is defined locally.  Just go to a basement with no windows and watch for change; you won't have to look so carefully as to be sensitive to global space-time geometry.

The nexus of the problem of time, as identified by Belot, is related to the  space on which one formulates the Hamiltonian dynamics.  ``This is the nexus of the problem of time:  time is not represented in general relativity by a flow on a symplectic space and change is not represented by functions on a space of instantaneous or global states.'' \cite[p. 209 ]{BelotPoP}.  But one ought not to try to represent time and change in GR in phase space; one needs phase space extended by a time coordinate because the gauge transformations are velocity-dependent   \cite{MukundaSamuelConstrainedGeometric,SuganoGaugeGenerator,SuganoGeneratorQM,LusannaVelocityHamiltonian}.  
To express 4-dimensional coordinate transformations, one also needs the lapse and the shift vector.  If one wishes to generate 4-dimensional coordinate transformations \emph{via} a Poisson bracket, then one needs a big Poisson bracket including the canonical momenta conjugate to the lapse and shift---the momenta that vanish according to the primary constraints.  
As noted above, it isn't terribly clear what has become of the lapse and the shift vector in Belot's treatment;  the quantities that he (like many authors) mostly discusses are  merely the 3-metric and its canonical momentum, giving 12 functions at each point in space, satisfying 4 constraints $\mathcal{H}_0=0$ and $\mathcal{H}_i=0$ at each point in space.  When one restores the lapse and shift vector to include the whole space-time metric and hence a better shot at expressing coordinate transformations involving time, and restores their conjugate momenta (vanishing as primary constraints) to improve one's chances at expressing 4-dimensional coordinate transformations using Hamiltonian resources, namely the gauge generator $G$, one gets 
20 functions of time at each point in space, satisfying 8 constraints at each point in space: $p=0$, $p_i=0,$ $\mathcal{H}_0=0$, and $\mathcal{H}_i=0$ at each point in space.  But to include velocity-dependent gauge transformations (such as change the time slice in GR), one should also include time in an extended phase space.  Hence instead of  $12 \infty^3$ functions 
satisfying $4 \infty^3$ constraints and (maybe) changing over a time that isn't part of the space in question, one needs $20 \infty^3$ quantities satisfying $8 \infty^3$ constraints 
on a slice of a space of $20 \infty^3 + 1$ dimensions. (Admittedly the primary constraints $p,$ $p_i$ are $4\infty^3$ quantities with the boring task of being $0$ according to $4 \infty^3$ of the constraints.) Such a space admits a Hamiltonian formalism equivalent to the Lagrangian formalism and hence makes change (or its absence) unproblematic in terms of the absence (or presence, respectively) of a time-like Killing vector field.   
Change apparently is not easy to find using reduction (at least, not using reductions attempted thus far). But it is readily found using the Lagrangian-equivalent non-reduced phase space including the primary constraints and extended by time, to which one can apply a gauge-invariant test for the existence of a gauge (coordinate system, in vacuum GR) in which change is absent.


\subsection{Th\'{e}bault on Time, Change and Gauge in GR and Elsewhere}

Unusually among philosophers, Th\'{e}bault's work has expressed at least selective skepticism about whether a first-class constraint generates a gauge transformation, especially in relation to time and the Hamiltonian constraint in theories or formulations that have one \cite{ThebaultReductionProblemofTime,ThebaultCanonicalGRTime}.  Such skepticism is partly informed by, among other things, views of Kucha\v{r}, Barbour and Foster's work, and of the Lagrangian equivalence-oriented reforms of Pons,  Salisbury and Shepley.  Clearly such skepticism could only be bolstered by the recognition  that a first-class constraint \emph{typically} does not generate a gauge transformation.  Then the quick argument from the fact that the Hamiltonian of GR is a sum of first-class constraints (and a boundary term) to the conclusion that it generates just a pile of gauge transformations is no longer tempting.  When familiar general presumptions about gauge freedom and first-class constraints disappear, less work is required to motivate taking GR as partly violating those presumptive conclusions. For example, Kucha\v{r}'s and Th\'{e}bault's exceptional treatment of the Hamiltonian constraint \emph{vis-a-vis} the other constraints in GR becomes partly unnecessary, because the other constraints are no longer viewed as having some of the features that Kucha\v{r} \emph{et al.} deny of the Hamiltonian constraint. (Of course the considerations about reduced phase space-time and the merely on-shell nature of Hamiltonian coordinate transformations due to the quadratic-in-momenta character of the Hamiltonian constraint imply that there are still some exceptional features of $\mathcal{H}_0$, rightly highlighted by Th\'{e}bault's doubts about reduced phase space.) The dual role of the Hamiltonian constraint in relation to both evolution and gauge transformations \cite{ThebaultCanonicalGRTime} is clarified when one notices that $\mathcal{H}_0$ does \emph{neither} of these jobs by itself; both are accomplished by teaming up with other constraints, whether in $H_p$ or in $G$. These teaming arrangements are easy enough to see when one retains the lapse, shift vector, and associated canonical momenta $p,$ $p_i$ and associated primary constraints, but impossible to see when one truncates the phase space in the way common since Dirac \cite{DiracHamGR,SalisburyBergmann,SalisburyRosenfeldMed}.   The idea of extending  phase space by $t$ in order to accommodate velocity-dependent gauge transformations also fits  well with Th\'{e}bault's project.  In short, making the constrained Hamiltonian formalism equivalent to the Lagrangian formalism as far as possible will facilitate drawing various conclusions for which Th\'{e}bault has argued on partly different grounds.


\section{ Quantum Problem of Time from Constraint Imposition }

While  change in Hamiltonian General Relativity is unproblematic at the classical level, the same does not hold  at the quantum level.  In other words, what remains of the problem of time, what actually exists of the problem, begins at quantization. It is many-faceted  \cite{ButterfieldIshamArguments,ButterfieldIsham,AndersonTimeAnnalen}. 
The usual Dirac method of imposition of constraints by requiring that a physical state be annihilated by them seems to close the door to change at the quantum level in a way with no classical analog \cite{PonsDirac}. 

%
%


\section{Conclusion}

  Having delved into the Hamiltonian formalism, one finds  that, when set up properly, its verdict on change agrees with that of  the condition of having no time-like Killing vector. Earman's healthy respect for a constrained Hamiltonian formalism and Maudlin's healthy respect for common sense and the superiority of the Lagrangian formalism 
are reconciled. 
This is of course no accident: making the Hamiltonian match the Lagrangian has been the basic policy of reform employed by Pons, Salisbury and Shepley's series of works, which I quote again:
\begin{quote}
We have been guided by the principle that the Lagrangian and Hamiltonian formalisms should be equivalent{\ldots}in coming to the conclusion that they in fact are.  \cite[p. 17]{PonsReduce} 
\end{quote}
Enforcing the equivalence of the unclear Hamiltonian formalism with the clear Lagrangian formalism is very evident in the early work of Bergmann's school---\emph{e.g.}, \cite{AndersonBergmann}.  But eventually, perhaps by accident, certain 
shortcuts were taken, by Bergmann about observables, by Dirac about gauge transformations, by both about whether the lapse, shift vector, and their canonical momenta should be retained in the phase space (\emph{e.g.}, \cite{DiracHamGR})---shortcuts yielding `insights' that have sustained confusion for decades.  Recovering Hamiltonian-Lagrangian equivalence has been underway for some time \cite{MukundaGaugeGenerator,CastellaniGaugeGenerator,SalisburySundermeyerEinstein,SuganoGaugeGenerator,GraciaPons}. While the gauge generator by now is moderately famous again, it has by no means swept the field.  Furthermore, and more importantly for present purposes, there remains the need to clear away the conceptual errors generated during the gauge generator's period of eclipse.  This paper has aimed to do that regarding time evolution.  A successor paper is planned to do the same regarding observables.

\section{Acknowledgements}

I thank Jeremy Butterfield for helpful discussion and comments on the manuscript, George Ellis,  Josep Pons, Claus Kiefer, Kurt Sundermeyer, David Sloan, Karim Th\'{e}bault, Oliver Pooley, and Edward Anderson  for helpful discussion, Jim Weatherall for helpful discussion of the hole argument and G. E. Moore,  
 and audiences in Cambridge, Oxford, Munich, Chicago, London (Ontario), and London. 



\end{document}